\documentclass[longauth]{aa}
\usepackage {epsfig,graphicx,color}
\usepackage {txfonts  }
\usepackage {natbib   }
\usepackage {float    }
\usepackage {enumerate}
\usepackage {tabularx }
\usepackage {url      }
\usepackage {capt-of  }

\begin{document}

%


\title {Pillars and globules at the edges of \ion{H}{ii} regions} 

\subtitle{Confronting {\it Herschel}\thanks{{\it Herschel} is an ESA space observatory with science instruments
  provided by European-led Principal Investigator consortia and with
  important participation from NASA.} observations and numerical
  simulations
}

\author{ 
P. Tremblin        \inst{1,2}  \and 
V. Minier          \inst{1}    \and 
N. Schneider       \inst{3,4}  \and
E. Audit           \inst{1,5}  \and
T. Hill            \inst{1}    \and
P. Didelon         \inst{1}    \and
N. Peretto         \inst{6}    \and
D. Arzoumanian     \inst{7}    \and
F. Motte           \inst{1}    \and
A. Zavagno         \inst{8}    \and
S. Bontemps        \inst{3,4}  \and
L. D. Anderson     \inst{9}    \and
Ph. Andr\'e        \inst{1}    \and
J. P. Bernard      \inst{10}   \and
T. Csengeri        \inst{11}   \and
J. Di Francesco    \inst{12}   \and
D. Elia            \inst{13}   \and
M. Hennemann       \inst{1}    \and
V. K\"onyves       \inst{1,7} \and
A. P. Marston      \inst{14}   \and 
Q. Nguyen Luong    \inst{15}   \and  
A. Rivera-Ingraham \inst{16,17}\and  
H. Roussel         \inst{18}   \and
T. Sousbie         \inst{18}   \and
L. Spinoglio       \inst{19}   \and
G. J. White        \inst{20,21}\and
J. Williams        \inst{22}
       }

\institute{Laboratoire AIM Paris-Saclay (CEA/Irfu - Uni. Paris Diderot
  - CNRS/INSU), Centre d'\'etudes de Saclay,  91191 Gif-Sur-Yvette,
  France
  \and
  Astrophysics Group, University of Exeter, EX4 4QL Exeter, UK
  \and
  Univ. Bordeaux, LAB, UMR 5804, F-33270, Floirac, France
  \and 
  CNRS, LAB, UMR 5804, F-33270, Floirac, France
  \and
  Maison de la Simulation, CEA-CNRS-INRIA-UPS-UVSQ, USR 3441, Centre
  d’\'etude de Saclay, 91191 Gif-Sur-Yvette, France
  \and
  School of Physics and Astronomy, Cardiff University, Queens
  Buildings, The Parade, Cardiff CF24 3AA, UK 
  \and
  IAS, CNRS (UMR 8617), Universit\'e Paris-Sud, B\^atiment 121, 91400
  Orsay, France
  \and
  Aix Marseille Universit\'e, CNRS, LAM (Laboratoire d'Astrophysique de
  Marseille) UMR 7326, 13388, Marseille, France  
  \and
  Department of Physics, West Virginia University, Morgantown, WV 26506, USA
  \and
  Universit\'e de Toulouse, UPS, CESR, 9 avenue du Colonel Roche, CNRS, UMR 5187,
  31028 Toulouse Cedex 4, France
  \and
  Max-Planck-Institut f\"ur Radioastronomie, Auf dem H\"ugel 69, 53121 Bonn, Germany
  \and
  National Research Council of Canada, Herzberg Institute of
  Astrophysics, 5071 West Saanich Road, Victoria, BC V9E 2E7, Canada 
  \and
  INAF – IAPS, via Fosso del Cavaliere 100, 00133 Roma, Italy
  \and
  European Space Astronomy Centre, Urb. Villafranca del Castillo, PO
  Box 50727, E-28080 Madrid, Spain 
  \and
  Canadian Institute for Theoretical Astrophysics, University of
  Toronto, 60 St. George Street, Toronto, ON M5S 3H8, Canada 
  \and
  Universit\'e de Toulouse; UPS-OMP; IRAP;  Toulouse, France
  \and
  CNRS; IRAP; 9 Av. colonel Roche, BP 44346, F-31028 Toulouse cedex 4, France
  \and
  Institut d'Astrophysique de Paris, Universit\'e Pierre et Marie Curie
  (UPMC), CNRS (UMR 7095), 75014 Paris, France 
  \and
  APS-INAF, Fosso del Cavaliere 100, I-00133 Roma, Italy 
  \and
  The Rutherford Appleton Laboratory, Chilton, Didcot, OX11 0NL, UK
  \and
  Department of Physics and Astronomy, The Open University, Milton Keynes, UK
  \and 
  Institute for Astronomy, University of Hawaii, 96822
  Honolulu, Hawaii}

\date{\today}
\mail{pascal.tremblin@cea.fr}

\titlerunning{Observations of \ion{H}{ii}
  regions in the Rosette and Eagle Nebula}
\authorrunning{Tremblin et al.}

\abstract{$Herschel$ far-infrared imaging observations have revealed the density structure of the interface between \ion{H}{ii}
  regions and molecular clouds in great
  detail. In particular, pillars and globules are 
  present in many high-mass star-forming regions, such as the Eagle
  nebula (M16) and the Rosette molecular cloud, and understanding
  their origin will help characterize triggered star formation.}
{The formation mechanisms of these structures are still being debated. The
  initial morphology of the molecular cloud and its turbulent state
  are key parameters since they generate deformations and curvatures
  of the shell during the expansion of the \ion{H}{ii} region. Recent numerical
  simulations have shown how pillars can arise 
  from the collapse of the shell in on itself and how globules can be
  formed from the interplay of the turbulent molecular cloud
  and the ionization from massive stars. The goal here is to test this
  scenario through recent observations of two massive star-forming
  regions, M16 and the Rosette molecular cloud.
}
{First, the column density structure of the interface between
  molecular clouds and associated \ion{H}{ii} regions was characterized using
  column density maps obtained from far-infrared imaging of the $Herschel$ HOBYS
  key programme. Then, the DisPerSe algorithm was used on these maps to detect
  the compressed layers around the ionized gas and 
  pillars in different evolutionary states. Column density
  profiles were constructed. Finally, their
  velocity structure was
  investigated using CO data, and all observational signatures were
  tested against some distinct diagnostics established from
  simulations.
}
{The column density profiles have revealed the importance of compression at
  the edge of the ionized gas. The velocity properties of the
  structures, i.e. pillars and globules, are very close to what we
  predict from the numerical 
  simulations. We have identified a good candidate of a nascent pillar in the
  Rosette molecular cloud that presents the velocity pattern of the
  shell collapsing on itself, induced by a high local curvature.
  Globules have a bulk velocity dispersion that indicates the
  importance of the initial turbulence in their formation, as proposed
  from numerical simulations. Altogether, this study re-enforces the
  picture of pillar formation by shell collapse and globule formation
  by the ionization of highly turbulent clouds.}  {}


\maketitle


%
%

\begin{figure*}[t]
\centering
\includegraphics[width=0.49\linewidth]{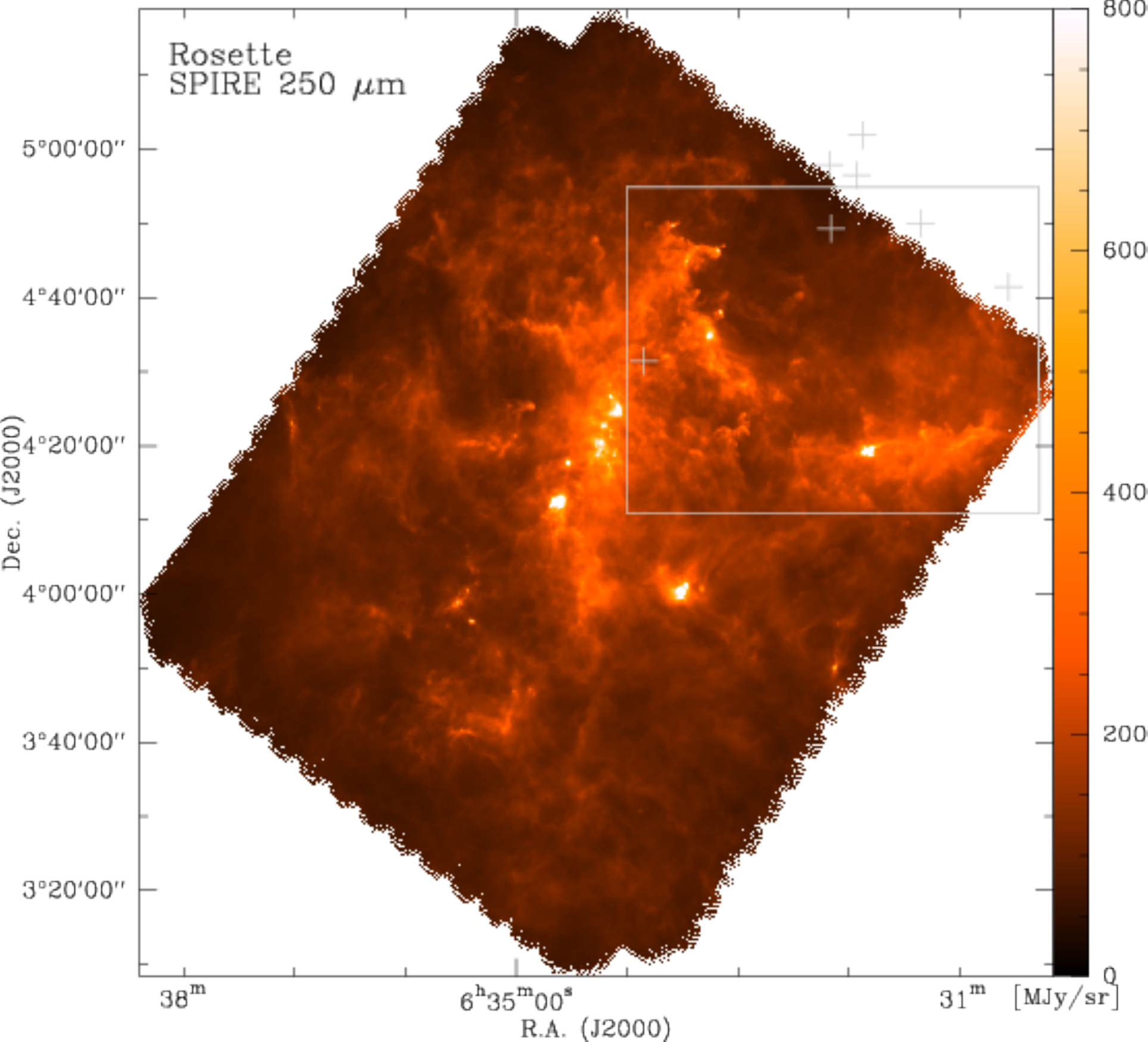}
\includegraphics[width=0.49\linewidth]{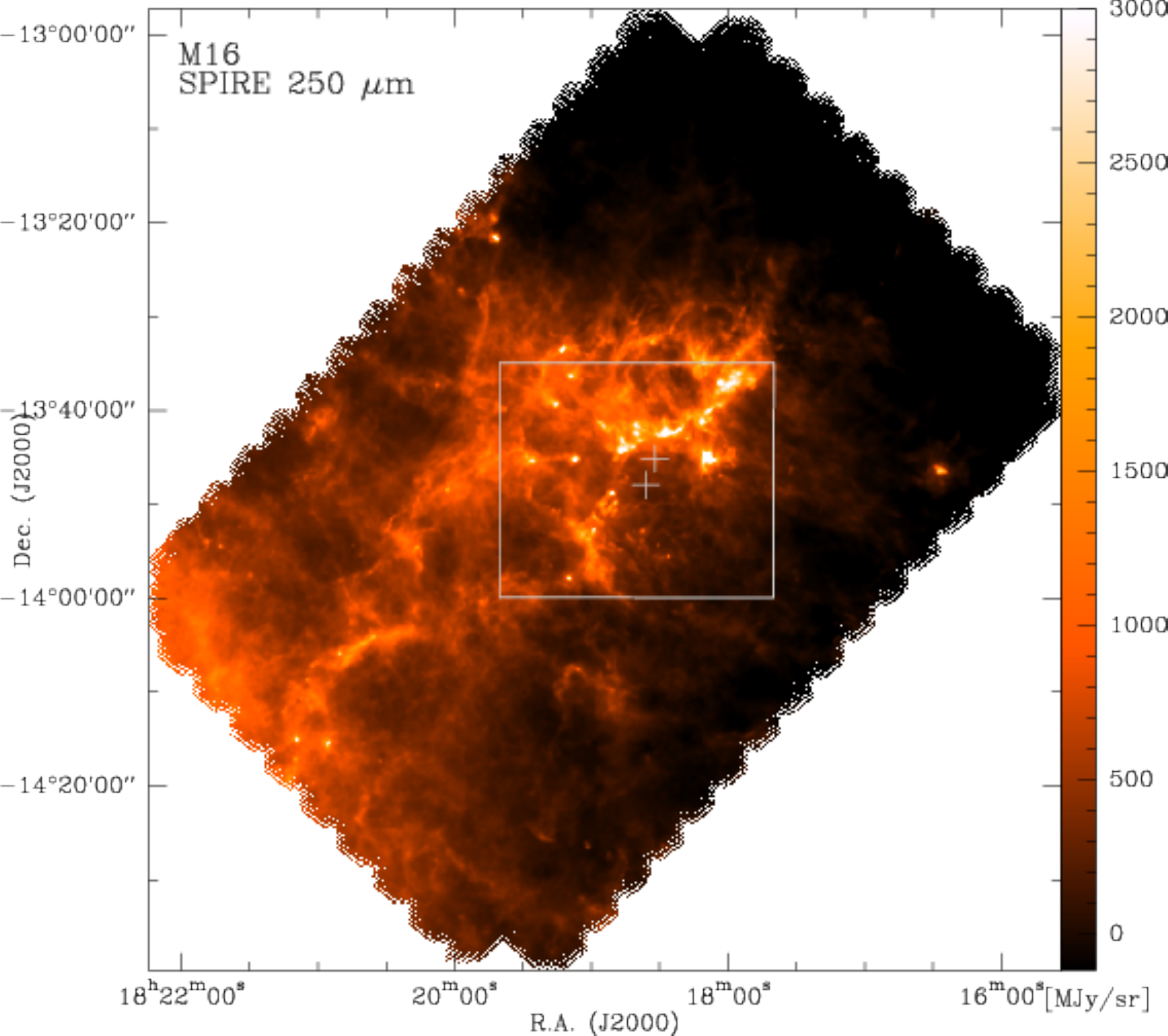}
\caption{\label{regions_over} {\sl Herschel} SPIRE 250 $\mu$m image of
  the Rosette (left) and M16 (right) molecular cloud with the studied areas outlined
  in grey (see Figs.~\ref{rosette_N} and \ref{rosette_rgb_spectra} for
  Rosette and Figs.~\ref{m16_N} and \ref{m16_rgb_spectra} for M16).
  The position of the main ionizing sources (O stars) are indicated
  with the grey crosses. The flux scale is in [MJy/sr].}
\end{figure*}

\section{Introduction}

\ion{H}{ii} regions can have very complex shapes depending on the
distribution of the ionizing sources and the initial structures in the
surrounding gas. They range from spherical bubbles \citep[e.g.][for
RCW120]{Deharveng:2009kd}, bipolar nebulae \citep[e.g. RCW
36][]{Minier:2013ih} to complex and large regions like Cygnus X OB2
and OB9
\citep[see][]{Schneider:2012hz}. Dense layers of gas and dust are
observed at the interface between 
the \ion{H}{ii} regions and their parent molecular cloud, 
and they consist of various structures. Condensations of cold gas are
often present in these layers
\citep[e.g.][]{Deharveng:2010cp}. Furthermore, \citet{Thompson:2012gn} shows that
massive young stellar objects (YSOs) are often
observed on the edge of the bubbles and that 15-30 \% of massive
stars could be formed in those regions.     
Elongated columns (pillars) of gas are observed pointing towards the
ionizing sources and are usually connected to the molecular cloud
\citep[e.g.][for "the Pillars of Creation"]{Hester:1996ir}. Unlike
pillars and condensations, cometary globules are bubbles of cold gas that are inside the
\ion{H}{ii} region and disconnected from the molecular cloud
\citep[e.g.][]{White:1986uy,White:1993ua,Thompson:2004ed,Schneider:2012hz}. Understanding the origin of these different structures is very
important for characterizing triggered star formation and the
impact of massive stars on the initial mass function (IMF).

Various scenarios have been investigated to explain the formation
mechanism of these structures. The collect and collapse scenario
\citep[see][]{Elmegreen:1977iq} proposes that the shell becomes
sufficiently dense thanks to cooling of the swept-up gas, to reach the
gravitational instability and therefore form dense condensations. This
scenario leads to a sequential star formation observed in many regions
\citep[e.g.][]{Blaauw:1991ux}. 
\citet{Bertoldi:1989bq} proposed that cometary globules could be the
result of the radiative implosion of an isolated
gravitationally stable clump. More recently, 
this scenario has been tested further using numerical simulations
\citep{Miao:2006bx,Miao:2009ds,Miao:2010gr,Bisbas:2011kg,Haworth:2011gv}. \citet{Mackey:2010cv}
shows that the shadowing effect of such clumps in the density field
could also form pillars. All these studies have shown how the
gravitational collapse of a stable cloud can lead to triggered star
formation. However, using cloud-scale simulations, \citet{Dale:2013da}
shows that it can be problematic to distinguish triggered from pre-existing
star formation. Also recently, parsec-scale simulations
have shown that most of the observed structures can be formed from
the interaction between the turbulence of the molecular cloud and
the ionization
\citep{Mellema:2006if,Arthur:2011ck,Gritschneder:2010du,Walch:2013vw}. However
it is often difficult to understand the precise mechanisms at the
origin of the different structures in these complex simulations.

\citet{Tremblin:2012ej} proposed that the important
parameter for forming pillars and clumps at the interface is
the degree of curvature of the dense shell perturbed by pre-existing
structures. A high curvature triggers the 
collapse of the shell into itself to form pillars, whereas low curvature
only triggers the formation of dense clumps that are accelerated with
the shell and remain in the shell. This scenario has since been tested for
turbulent media \citep[see][]{Tremblin:2012he} in which the level of
turbulence (compared to the pressure of the \ion{H}{ii} region) has
been identified as a key ingredient in the formation of cometary
globules. These simulations do not take the FUV
  photons and the subsequent PDRs and temperature gradients in the cold gas into account. Many observational diagnostics based on the density and
velocity structures of the compressed cold gas at the interface have been derived to
test these models:
\begin{itemize}
\item Shell fragments are compressed layers of gas formed by the expansion of
  the ionized gas, and they
  can be identified by an excess in the column-density maps.
\item A nascent pillar can be
  identified thanks to a red-shifted and blue-shifted component in the
  velocity spectrum of the parts of the shell that are locally
    collapsing. The velocity shift 
  between the two peaks should be twice the velocity of the globally expanding
  shell. An evolved pillar shows a single peak at the velocity of
  expansion of the shell. 
\item A globule formed from the interaction of the turbulence and the
  ionization has a line-of-sight velocity shifted at high velocities
  compared to those of nearby dense shell fragments and pillars.
\end{itemize} 
The present paper aims at scrutinizing these
diagnostics using {\it Herschel} observations of the Rosette
Nebula and the Eagle nebula (M16) taken as part of the HOBYS key
project\footnote{
The \emph{Herschel} imaging survey of OB Young Stellar objects (HOBYS) 
is a \emph{Herschel} key program. See \url{http://hobys-herschel.cea.fr}}
 \citep[Herschel imaging survey of OB Young Stellar
  objects,][]{Motte:2010fy}. These two regions (overview in
 Fig.~\ref{regions_over}) present a complex 
 interface with pillars, globules, and compressed layers. They have a similar
 total ionizing flux and therefore are perfect laboratories for testing the
 diagnostics derived from the simulations.  
We first investigate in Sects.~\ref{rosette} and \ref{m16} the density
structure of the interface between the molecular and ionized gas by
tracing the crests of the densest parts (named dense fronts hereafter)
of column density maps using the DisPerSe algorithm
\citep{Sousbie:2011ft}, and study their profiles.  Using molecular
gas tracers, we then study the bulk velocity of the pillars, clumps,
and globules that are present in these regions. Finally
(Sect.~\ref{discussion}), we compare these observations with the
curvature model proposed in \citet{Tremblin:2012ej} for the formation
of pillars and dense clumps, and the turbulent one proposed in
\citet{Tremblin:2012he} for the formation of globules.

\section{The Rosette nebula}\label{rosette}

\indent The Rosette molecular cloud is located at 1.6 kpc from the Sun
\citep{Williams:1994hla,Schneider:1998ta,Heyer:2006hv}. The central
cluster NGC 2244 is dominated by 17 OB stars that have a total
Lyman-$\alpha$ luminosity of 3.8$\times$10$^5$ L$_\odot$
\citep[see][]{Cox:1990ub}. Photon-dominated regions (PDRs) are
detected deep inside the cloud \citep[see][]{Schneider:1998vx}. The
infrared (IR) population of the molecular cloud was investigated in
near- and mid-IR surveys \citep[see][and references
therein]{Poulton:2008fp}. More recently, the Rosette\textcolor[rgb]{0,0,0}{} molecular cloud
was studied using far-IR {\it Herschel} observations
\citep[see][]{Schneider:2010ec,Schneider:2012ds,Hennemann:2010hp,Motte:2010fy,DiFrancesco:2010ef}. Figure~\ref{regions_over} provides
  as an overview an improved SPIRE 250 $\mu$m map of the Rosette cloud,
  reduced with the pipeline in
  HIPE10\footnote{\url{http://herschel.esac.esa.int/hipe}}, which includes a
  destriping module with baseline subtraction, correction for extended
  emission, and an absolute flux calibration 
  using the {\sc zeroPointCorrection} task in HIPE10 to obtain 
  the offset values from Planck. The PACS 70 and 160 maps were produced using 
  the HIPE pipeline and the Scanamorphos sotware package \citep{Roussel:2012wv}.

In Sect. \ref{rosette_cn} we investigate the H$_2$ column density
structure of the edge of the molecular cloud using the {\it Herschel}
map with an angular resolution of 37$''$ (0.26 pc). We checked
the width of the pillar in the north-east of the map using
the 70 $\mu$m data (resolution of 
$\sim6''$). The 70 $\mu$m can be used as a tracer of the warm dust
in the photo-dissociation region (PDR) and therefore as an indicator
of the width of the cold structures surrounded by the PDR. The
resulting width is similar, and the structures are therefore well
resolved at the edge of the cavity.
 In Sect.~\ref{rosette_v} we concentrate on the velocity structure of three
subregions of the interface: the pillar (east of the cavity), the
globules (also east), and the southern part of the dense front.

\begin{figure}[t]
\centering
\includegraphics[trim=0 3cm 2cm 0cm ,width=\linewidth]{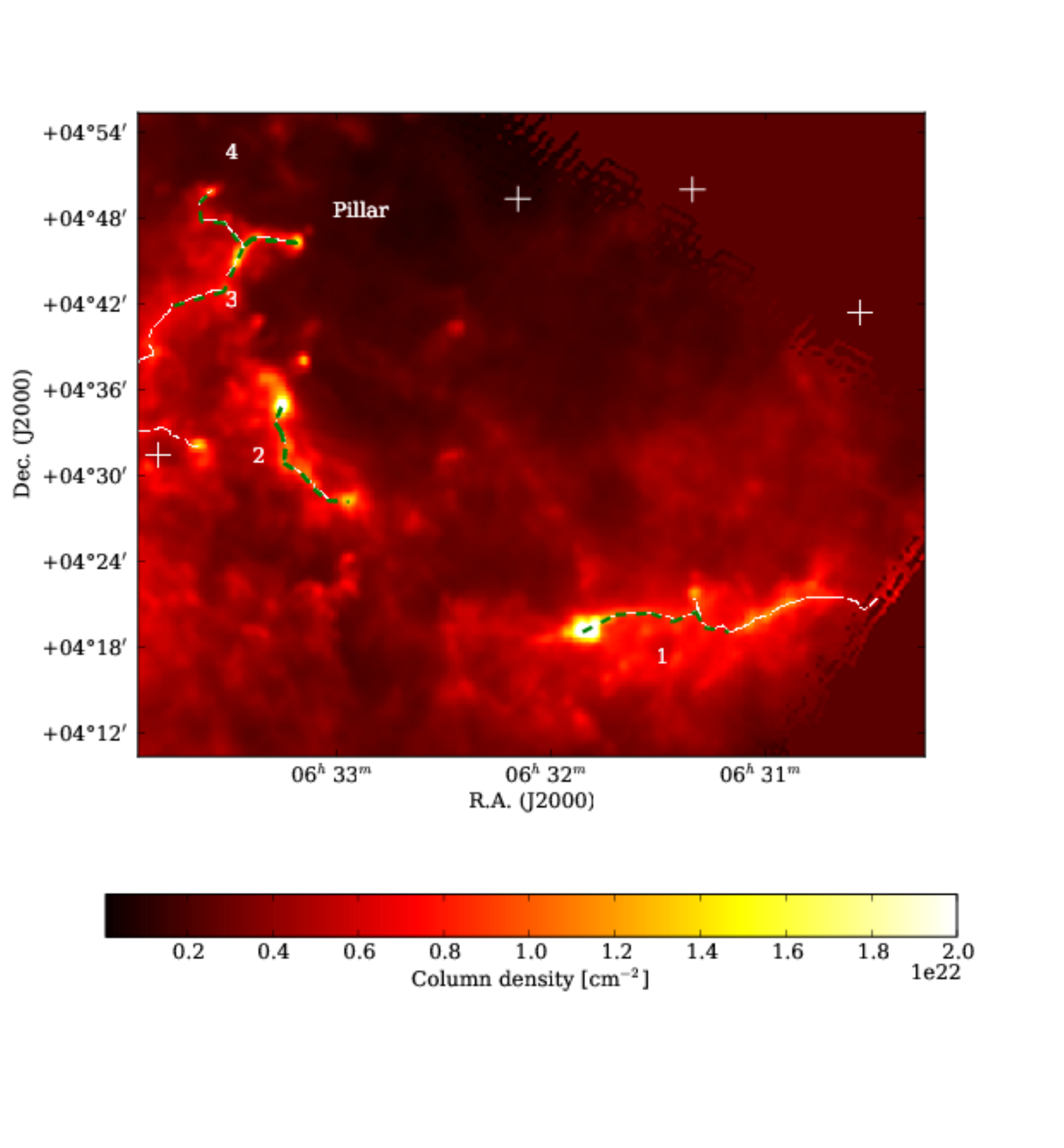}
\caption{\label{rosette_N} Column density map (with 37$''$ resolution) of the interface between
  the Rosette molecular cloud and the \ion{H}{ii} region around NGC
  2244. The
  DisPerSe skeleton is drawn in white, and each part of the dense
  layer is labelled from 1 to 4. The pillar is identified as a segment
  with one side
  connected by a triple point to the dense fronts 3 and 4, and the
  opposite side is disconnected and points toward the ionizing
  sources. The parts of the skeleton used for the 
  profiles in Fig.~\ref{rosette_shock_1} are indicated with the green
  dashed lines. The position of the main ionizing
  sources (O stars) are indicated with the white crosses. }
\end{figure}
\begin{figure}[t]
\centering
\includegraphics[trim=0 0 1.70cm 0
  ,width=0.49\linewidth]{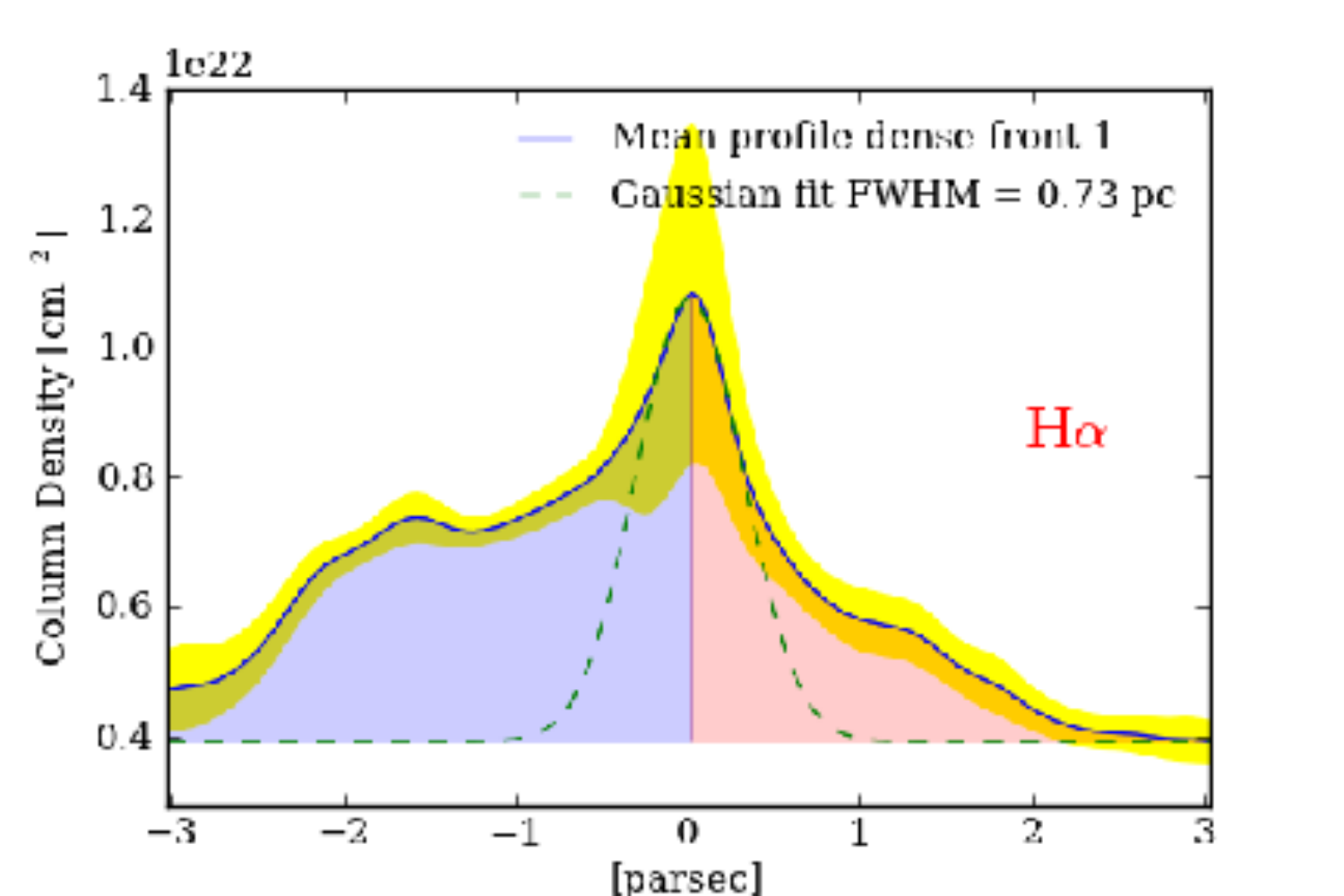} 
\includegraphics[trim=0 0 1.70cm 0
  ,width=0.49\linewidth]{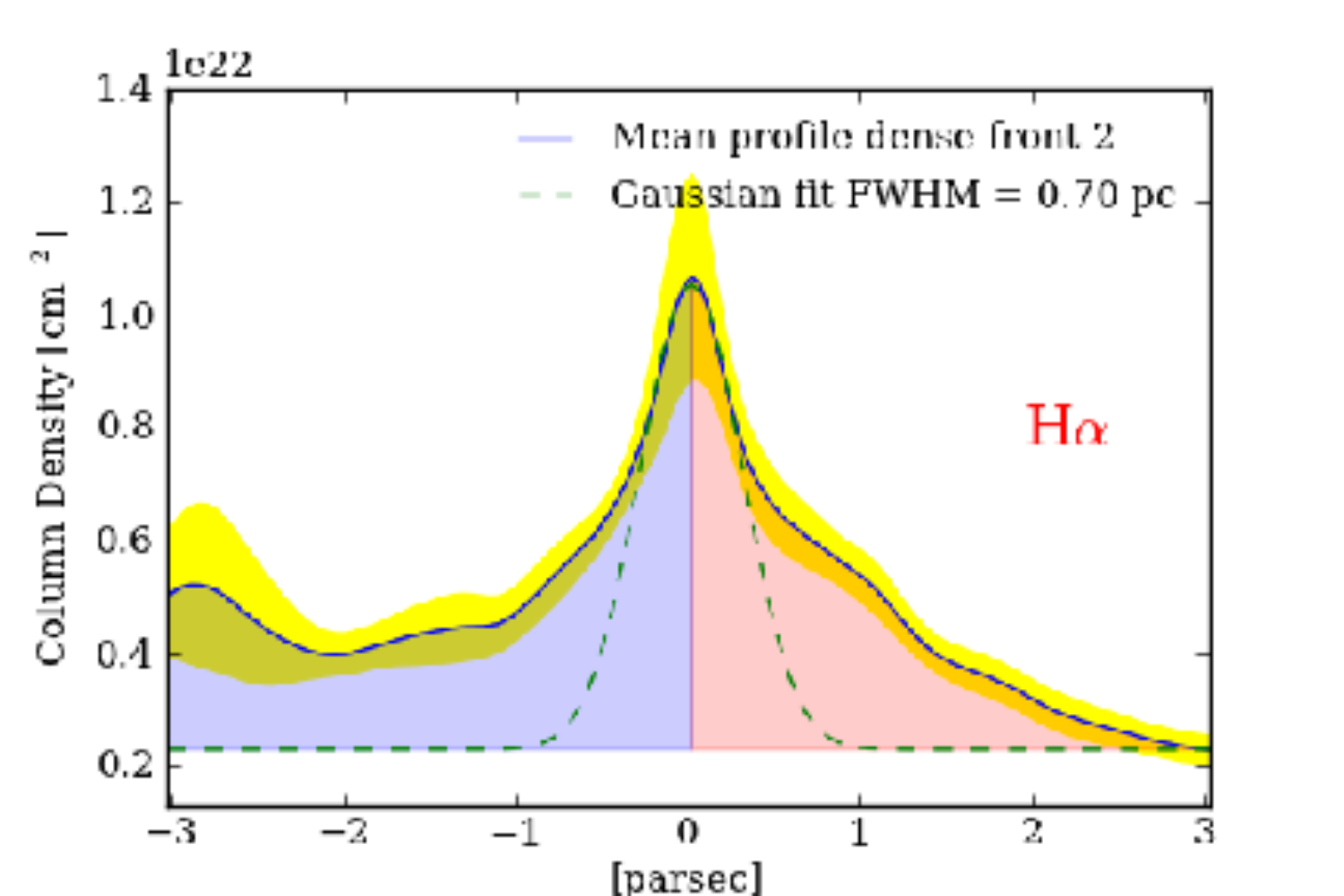}
\includegraphics[trim=0 0 1.70cm 0
  ,width=0.49\linewidth]{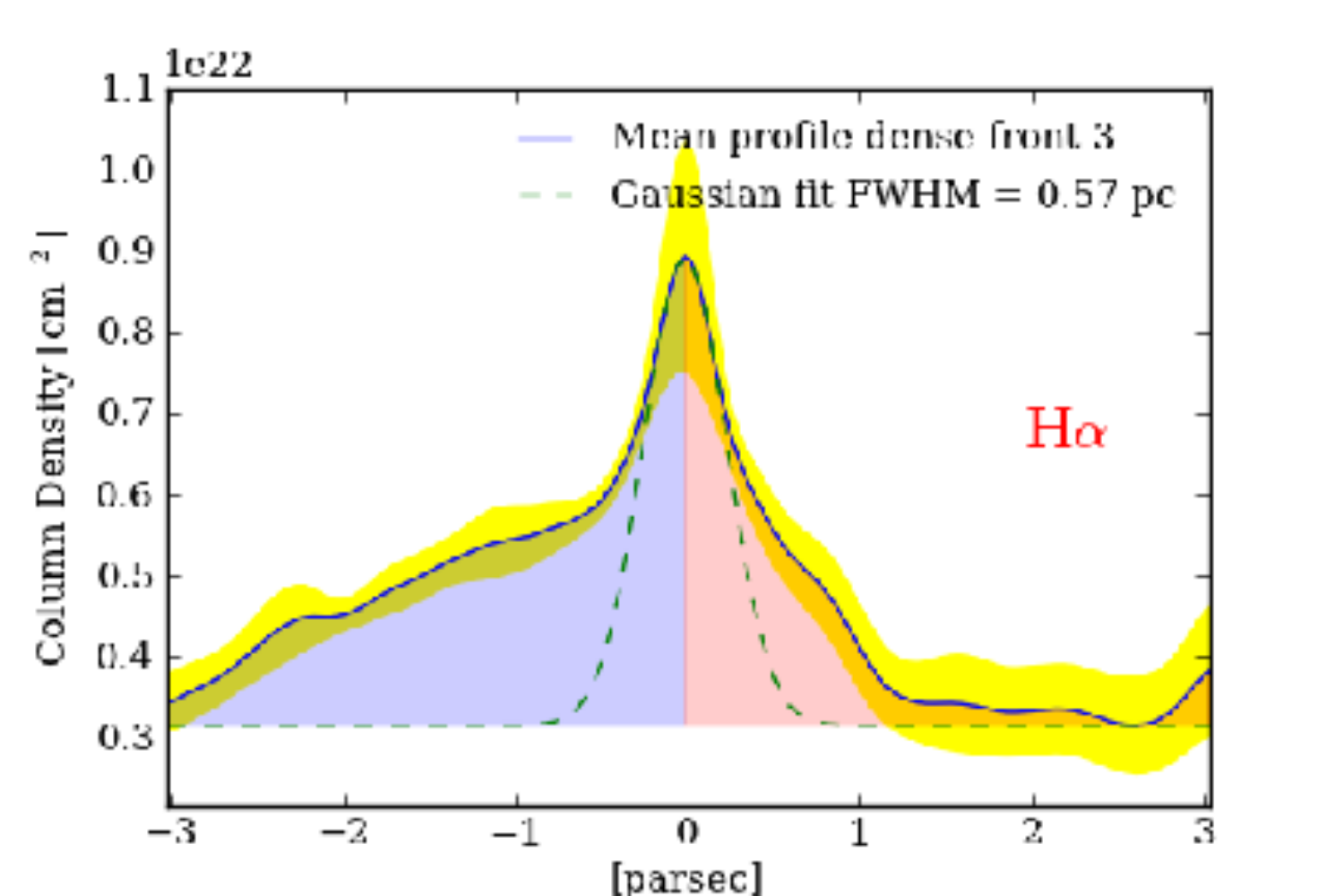} 
\includegraphics[trim=0 0 1.70cm 0
  ,width=0.49\linewidth]{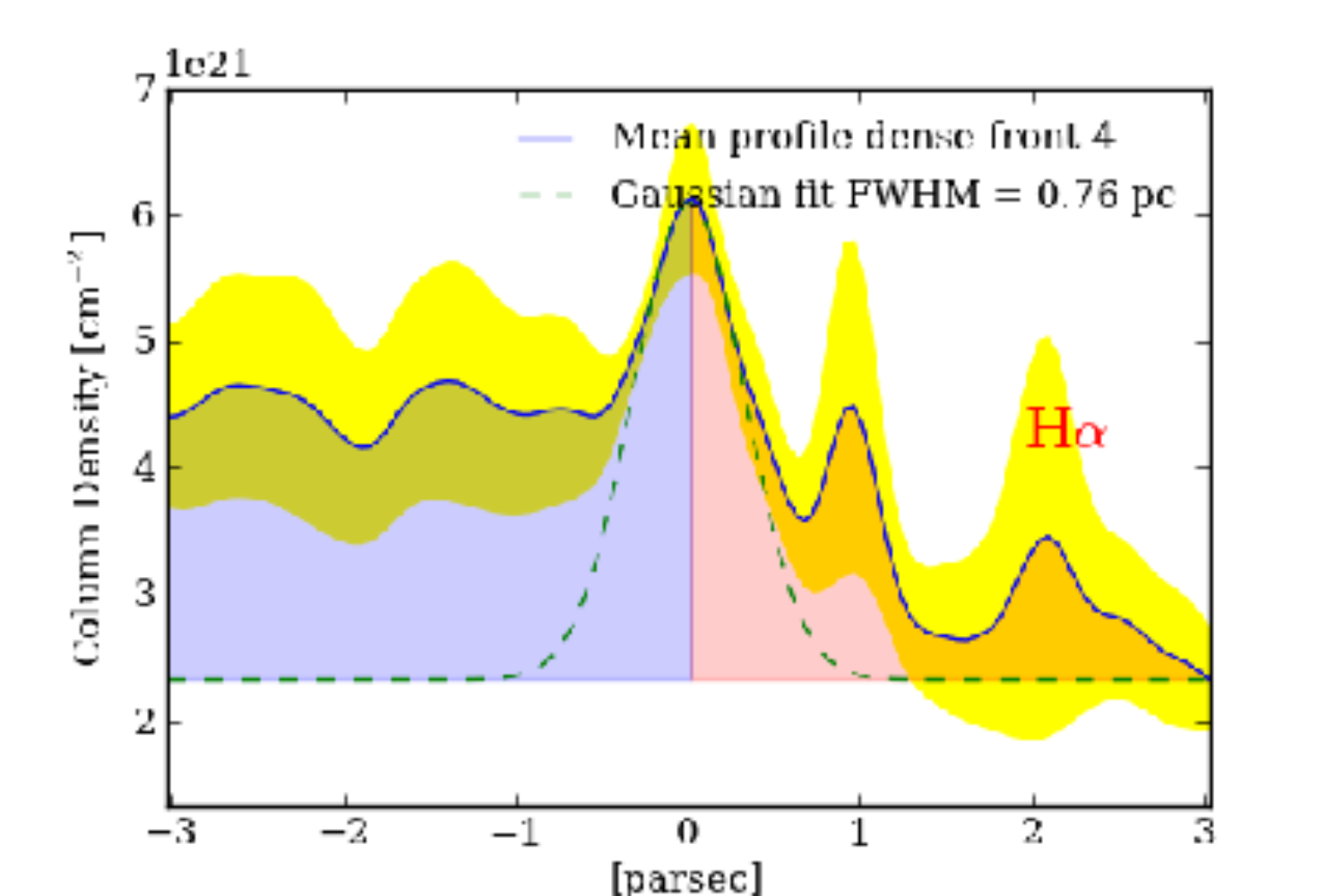} 
\includegraphics[trim=0 0 1.70cm 0
  ,width=0.49\linewidth]{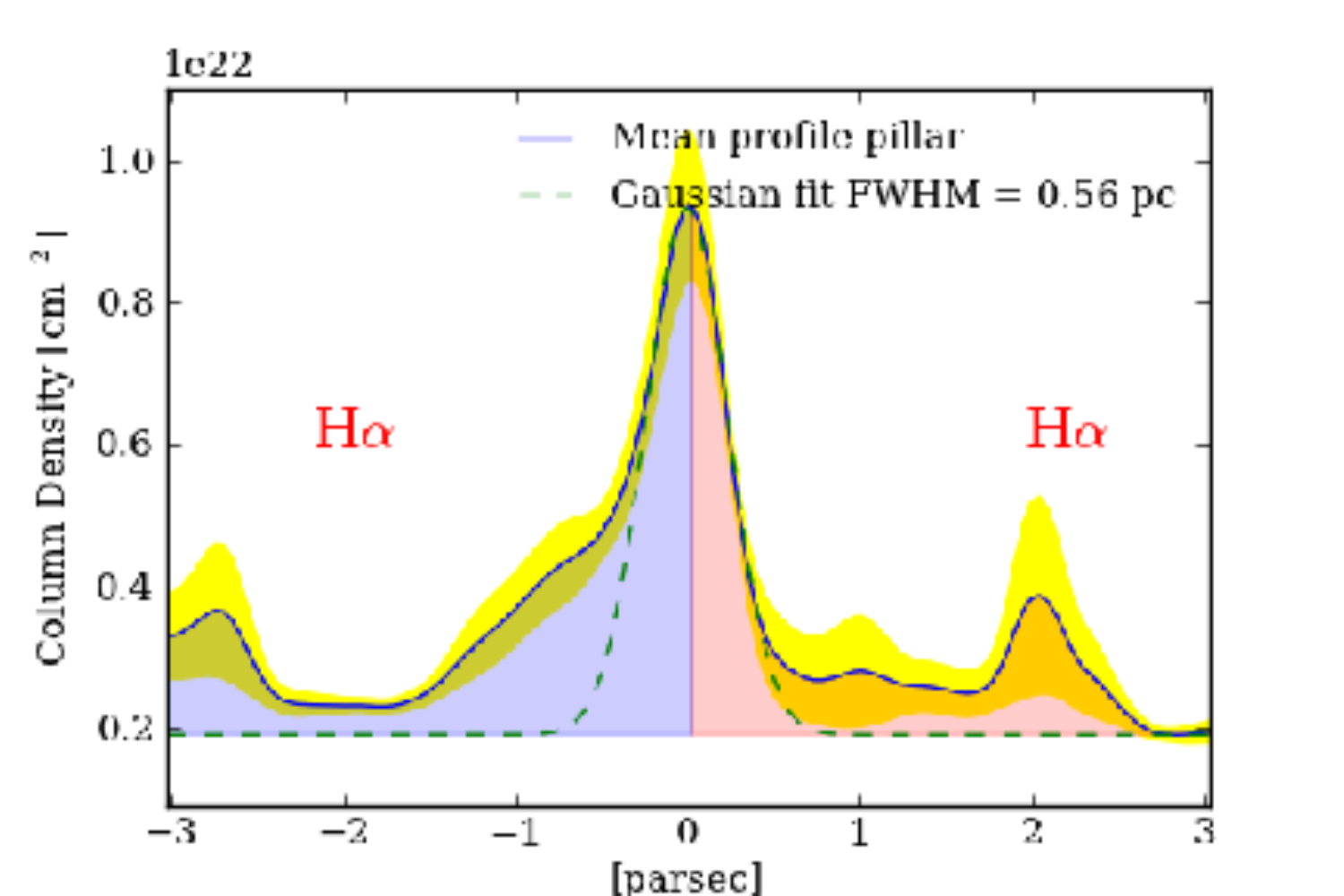}  
  \captionof{figure}{\label{rosette_shock_1} Profiles of the dense fronts and
    pillars traced by DisPerSe in the Rosette Molecular cloud. The
    extension of the profiles is fixed at 3 parsecs. The
    yellow shaded area represents the standard deviation of the
    profiles. The red sign H$\alpha$ indicates on which side ionized gas is
    present (see also Fig.~\ref{rosette_rgb_spectra} and the H$\alpha$
    emission). The green-dashed curve is the Gaussian fit used to
    determine the FWHM width.}
\textcolor{white}{.}\\
\centering
\begin{tabular}{l|cccc}
dense front & FWHM (pc) & $\sigma_{FWHM} (pc) $ & Asymmetry & $\sigma_{asym}$ \\
\hline
\hline
1      & 0.75 & 0.29 & 1.8 & 0.4 \\
2      & 0.69 & 0.20 & 1.4 & 0.6 \\
3      & 0.57 & 0.27 & 1.6 & 0.8 \\
4      & 0.76 & 0.23 & 2.2 & 1.4 \\
pillar & 0.55 & 0.05 & 1.8 & 1.0 \\
\end{tabular}
\captionof{table}{\label{rosette_param} Deconvolved FWHM width and asymmetric parameter for
  the averaged profiles of the 5 dense fronts in the Rosette Molecular
  cloud. The asymmetric parameter is the ratio of the blue area to the
red area in the profiles in Fig.~\ref{rosette_shock_1}.}
\end{figure}

\subsection{Column density structure}\label{rosette_cn}

The column density map of the Rosette cloud was made by fitting
  pixel-by-pixel the spectral energy distribution (SED) of a greybody
  to the {\sl Herschel} wavebands between 160 and 500 $\mu$m (at
  the same $37''$ resolution), assuming the dust opacity law of
  \citet{Hildebrand:1983tm} and an emissivity spectral index of
  $\beta=2$. The 70 $\mu$m 
  data are not used to derive the column density since it does not
  entirely trace the cold dust. See \citet{Schneider:2012ds} for more
  details.  The resulting column density map obtained from {\sl Herschel} data 
covers a wider density range and is thus more representative than
column density maps obtained from molecular line data because there is
no cut-off at high or low densities as  is the case for, say,
$^{13}$CO \citep[see][for a detailed comparison in W43]{Carlhoff:2013td}.
Figure~\ref{rosette_N} shows the column density map of the
  interface between the Rosette molecular cloud and the \ion{H}{ii}
  region around NGC 2244. We then applied the DisPerSe algorithm
  \citep{Sousbie:2011ft}: a general method, based on principles of
  computational topology that traces filaments by connecting saddle
  points to maxima with integral lines. DisPerSe was used with an
  intensity contrast level of 4$\times$10$^{21}$ cm$^{-2}$ and a
  low-density threshold of 5$\times$10$^{21}$ cm$^{-2}$. These
  parameters were chosen to trace only the crests of the main parts of
  the dense front that can be identified by eye on the column density
  map. As outlined in \citet{Schneider:2012ds}, a lower density contrast
  (e.g.  $\sim$0.5$\times$10$^{21}$ cm$^{-2}$) results in a very
  filamentary structure, while higher values lead to identifying
  of only the main features.

The dense fronts are labelled from 1 to 4 in Fig.~\ref{rosette_N}. A recent
  survey of far-IR clumps in the Rosette cloud (White et al. in
  prep.) shows that the bright far-IR source at the left-hand
  edge of dense front 1 shows H$_2$O maser emission,
  an observational signature that indicates star formation.
The pillar and the dense fronts
3 and 4 are connected, forming a triple point (the connection between
three lines) in the skeleton. Another
triple point is seen in the middle of the dense front 1.  We performed
transverse column density profiles along a piecewise linear
segmentation of the DisPerSe skeleton following a similar procedure to the one, say, in
  \citet{Arzoumanian:2011ho}. Figure~\ref{rosette_shock_1} presents the
averaged profiles on the dense fronts and on the pillar. 

\begin{figure}[t]
\centering
\includegraphics[trim=0 1cm 2cm 0cm ,width=\linewidth]{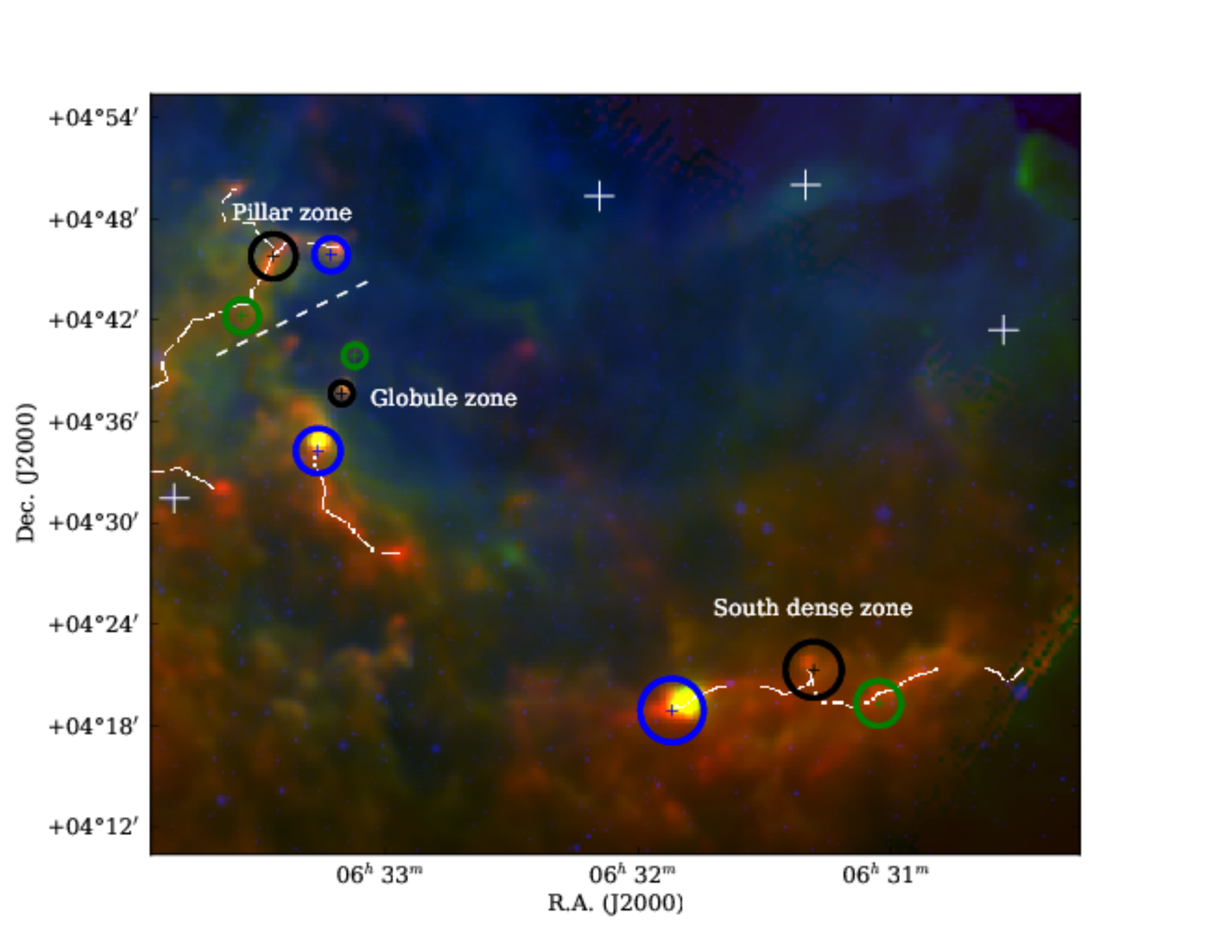}
\caption{\label{rosette_rgb_spectra} Three-color image of the interface between
  the Rosette molecular cloud and the \ion{H}{ii} region around NGC 2244 (red:
  {\it Herschel} column density map, green: PACS 70 $\mu$m, blue: H$\alpha$). The
  white crosses indicate the O stars. Each spectrum in
  Fig.~\ref{rosette_velocity} is integrated inside the corresponding
  colored circles. The centre and radius of each circle are chosen
  by taking the peak value and the extension of the CO emission. The
  white dashed line is an indication for the separation between the
  pillar and globule zones.} 
\end{figure}

\begin{figure}[t]
\centering
\includegraphics[trim=0 0 0cm 0
  ,width=0.8\linewidth]{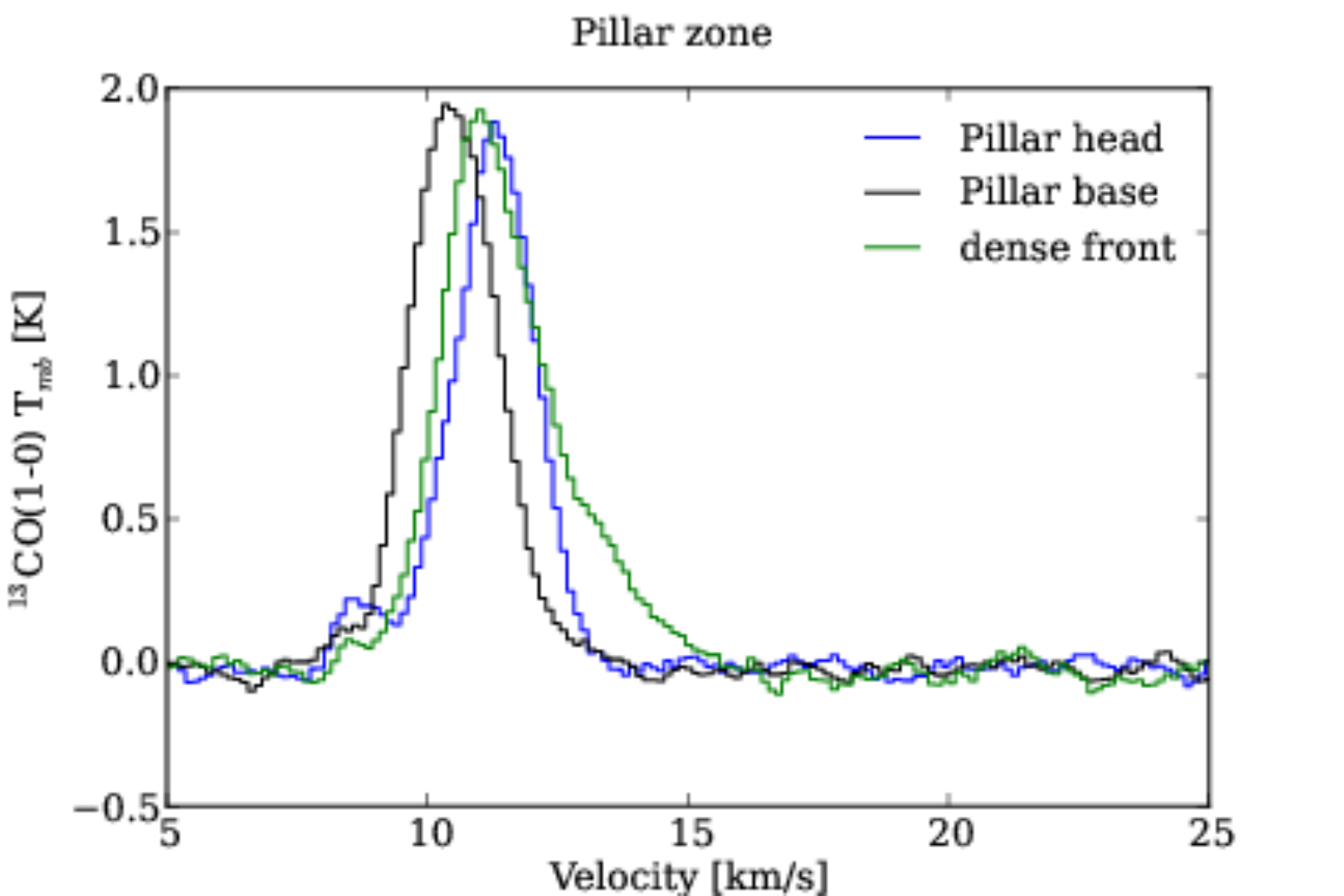} 
\includegraphics[trim=0 0 0cm 0
  ,width=0.8\linewidth]{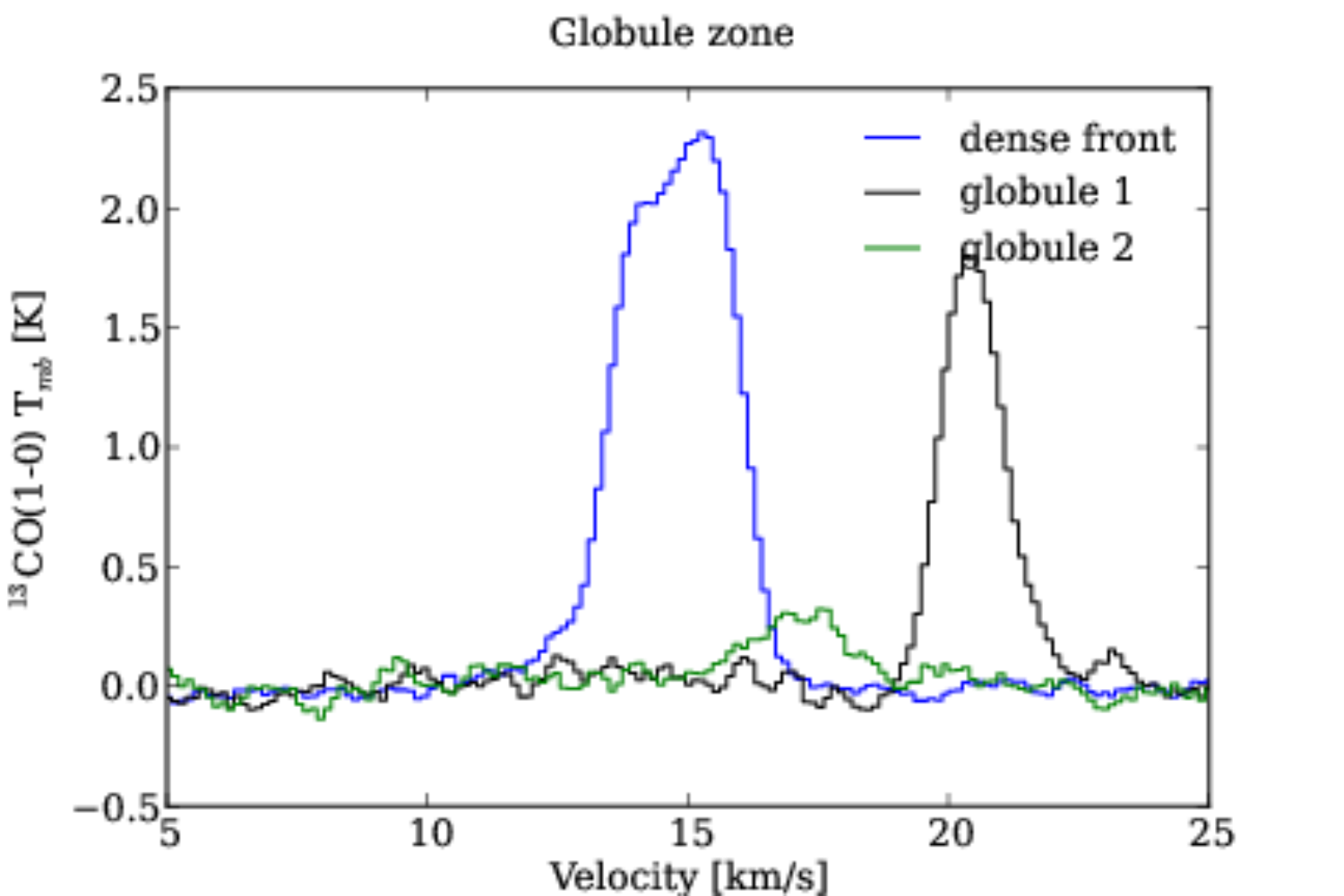}
\includegraphics[trim=0 0 0cm 0
  ,width=0.8\linewidth]{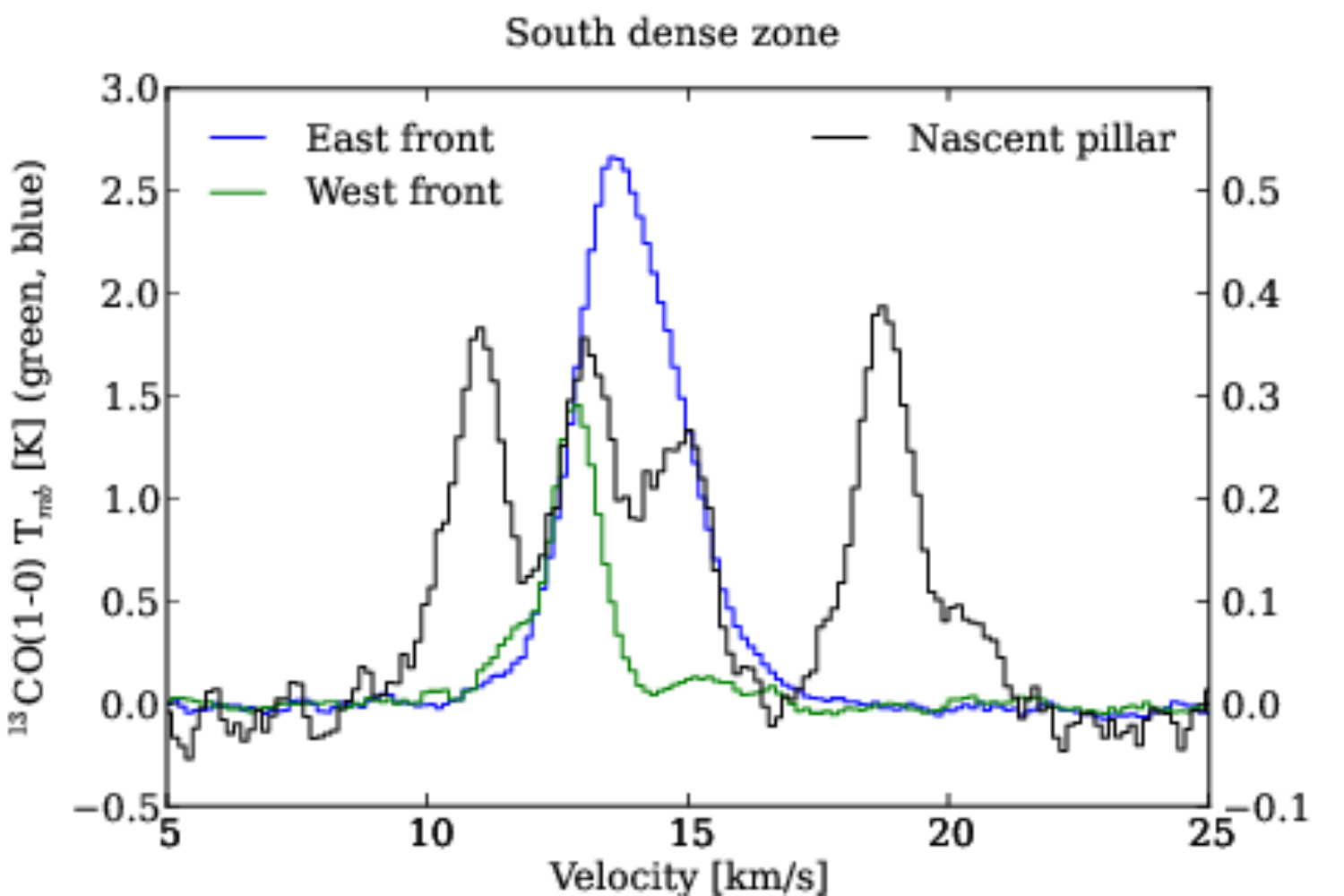} 
\caption{\label{rosette_velocity} $^{13}$CO (1-0) integrated
spectra in the three sub-regions of the interface in the Rosette molecular cloud. 
The spectra are spatially integrated in the colored circles showed in
Fig.~\ref{rosette_rgb_spectra}. The temperature scale for the nascent
pillar area on the South dense zone is multiplied by a factor of five.} 
\end{figure}

\begin{figure}[t]
\centering
\includegraphics[trim=0 0 0cm 0
  ,width=0.8\linewidth]{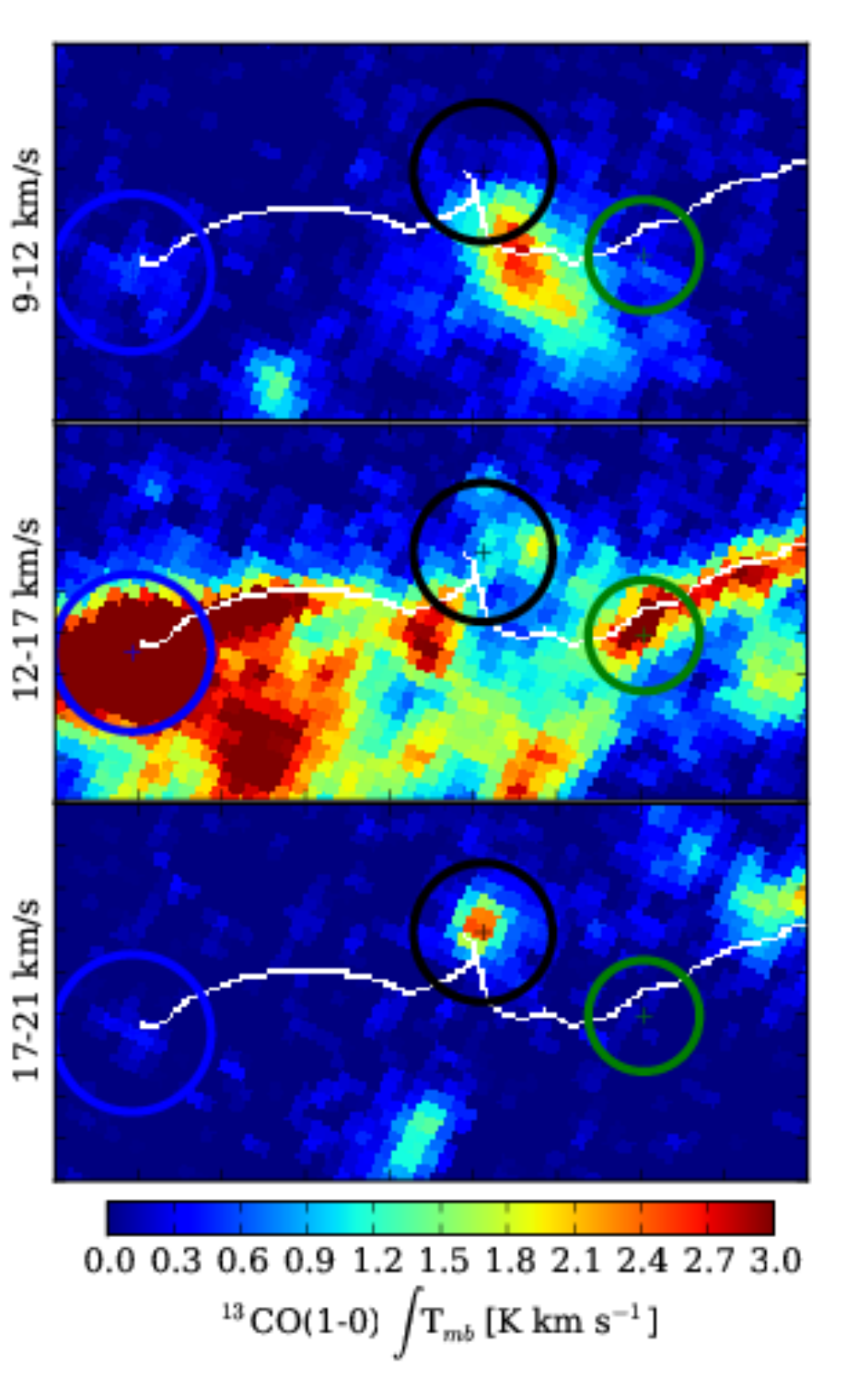} 
\caption{\label{channel_maps} $^{13}$CO (1-0) integrated channel maps of the
  south dense zone around the three main velocity components:
  9-12~km/s, 12-17~km/s, and 17-21~km/s.}  
\end{figure}

We fitted the inner part of the dense fronts with Gaussian functions
after subtracting a background set at the minimum of the fitted
profile. The level of the background does affect the width of the
Gaussian profile, however other choices of the background level only
induce small variations in the width as also found in \citet{Peretto:2012bu}. The
definition used here leads to a background around the same level in
all of the profiles considered N$_{H_2} \approx 2-4\times10^{21}$~cm$^{-2}$. This approach is also similar to the work
of \citet{Arzoumanian:2011ho} and \citet{Hennemann:2012dp}. There are two possibilities for
determing the full width at half maximum (FWHM): fitting Gaussian
profiles at each position along the front and taking the mean value or
fitting a Gaussian profile on the averaged profiles in Fig.
\ref{rosette_shock_1}. We used both strategies, and the FWHM
values along the fronts are given in Table~\ref{rosette_param}
(deconvolved with the beam of the telescope). The average and standard
deviation of the width along the fronts is weighted by the inverse of
the standard deviation of the fit at each position along the front.

The computed widths are in good agreement with the widths of the
average profiles given in Fig.~\ref{rosette_shock_1}. For example, the
averaged values of the deconvolved FWHM are 0.75 $\pm$ 0.3 pc for the
dense front 1 and 0.55 $\pm$ 0.05 pc for the pillar.  These values are
3 \% different from the width of the averaged profile in Fig.
\ref{rosette_shock_1} so they fit within the error bars.  The widths of the
pillar and the dense front 3 are smaller than values for other
parts of the front. Since these two structures are parallel to the
direction of the expansion of the \ion{H}{ii} region, and the other dense fronts are
perpendicular, a possible explanation is that the perpendicular dense
fronts accumulate more gas during the expansion, while this is not
possible for elongated structures parallel to the direction of
expansion. 

Similar to
\citet{Peretto:2012bu}, we defined an asymmetric parameter as the
ratio of the integrated column density on either side of the profile
peak (background subtracted). It corresponds to the blue area over the red area in Fig.
\ref{rosette_shock_1}, and the
values for the different profiles are 
indicated in Table~\ref{rosette_param}. The dispersion is the standard
deviation of the parameter value along the fronts. The dense fronts
are systematically asymmetric with less gas on the side that is exposed
to ionized gas (indicated by the red H$\alpha$ in
Fig.~\ref{rosette_shock_1}). The profile of the pillar is also
asymmetric, however we see in Sect.~\ref{m16} that there is no
preferential side for the asymmetry in a pillar profile.

\subsection{Velocity structure}\label{rosette_v}

The bulk velocity of the objects in the interface was investigated
using a $^{13}$CO (1-0) map obtained with the FCRAO\footnote{Five College Radio Astronomy Observatory} \citep[see][]{Heyer:2006hv}, with an
angular resolution of $45''$ at 115 GHz. We concentrated our study on
the bulk velocity of the main components of the spectra that will be
used to compare with models in Sect. \ref{discussion}. The three
different regions considered are named pillar, globule, and southern dense
zones hereafter, and they are indicated in
Fig.~\ref{rosette_rgb_spectra}. The areas in which integrated spectra
are indicated in the three-color image in Fig.
\ref{rosette_rgb_spectra} and the corresponding spectra are presented
in Fig.~\ref{rosette_velocity}. 

In the {\bf pillar zone}, the blue spectrum is integrated around the head of
the pillar, the black one around the triple point connecting the
pillar to the dense fronts 4 and 5, and the green one is integrated on a
part of the dense front 5. The main components of the spectra are
centred on the same velocity of 10-11 km s$^{-1}$. Dense
front 4 is also connected to the pillar and has a similar
velocity. Therefore, the pillar 
and the connected fronts have a velocity similar to the
expansion velocity of the bubble.

The blue spectrum in the {\bf globule zone} is integrated around a
clump in the dense front 2, and the black and green ones are integrated
around the globules 1 and 2 north of the front. Dense front 2 is
a nearby dense region associated with the molecular cloud. In
\citet{Schneider:1998ta}, a high angular resolution H$\alpha$ image
was shown with an overlay of $^{13}$CO (2-1) emission (their
    Fig.~15).  It clearly shows that the two globules are illuminated by
    the OB cluster, similar to what is seen at the edge of the dense front. Thus,
    despite the higher velocity of the globule 1 ($\sim$21 km s$^{-1}$)
    with respect to the bulk emission of the dense front of
    15~km~s$^{-1}$, the globule is close to the dense front.
  The globules
  have velocities red-shifted by +3 and +6 km s$^{-1}$ relative to the
  nearby front, whereas the pillar has the same velocity as the
  connected fronts. Therefore the pillar follows the expansion of the
  bubble, while the globules have motions that are not related to the
  direction of expansion. 

  In the {\bf south dense zone}, the blue and green spectra indicate
  the mean velocity of the western and eastern parts of the dense front,
  respectively. The western part has a velocity in the range 14-15 km
  s$^{-1}$, the eastern part of the order of 13 km s$^{-1}$. These two
  velocity components join at the triple point, which was previously
  identified in the DisPerSe skeleton (see Fig.~\ref{rosette_N}) where
  the black spectrum is integrated (see
  Fig.~\ref{rosette_rgb_spectra}). Moreover, there are two other 
  velocity components in the black spectrum, i.e. at 11 km s$^{-1}$
  and 19 km s$^{-1}$. This complex velocity structure
  with several components also becomes obvious in channel maps over  
the three main velocity ranges shown in Fig.~\ref{channel_maps}. The
bulk emission of 
the cloud at $\sim$12-17 km s$^{-1}$  
mainly corresponds to the column density map obtained from {\it Herschel},
and the DisPerSe filament tracing  
identifies this component. South of the triple point marked black,
however, we identify a 
gas reservoir at lower velocities, and at the northern ending tip of
the DisPerSe  
skeleton there is a smaller region at higher velocities. This characteristic
velocity distribution fits into theoretical  
predictions for pillar formation and is discussed in more detail
in Sect.~4.

\begin{figure}[t]
\centering
\includegraphics[trim=0 3cm 2cm 0cm ,width=\linewidth]{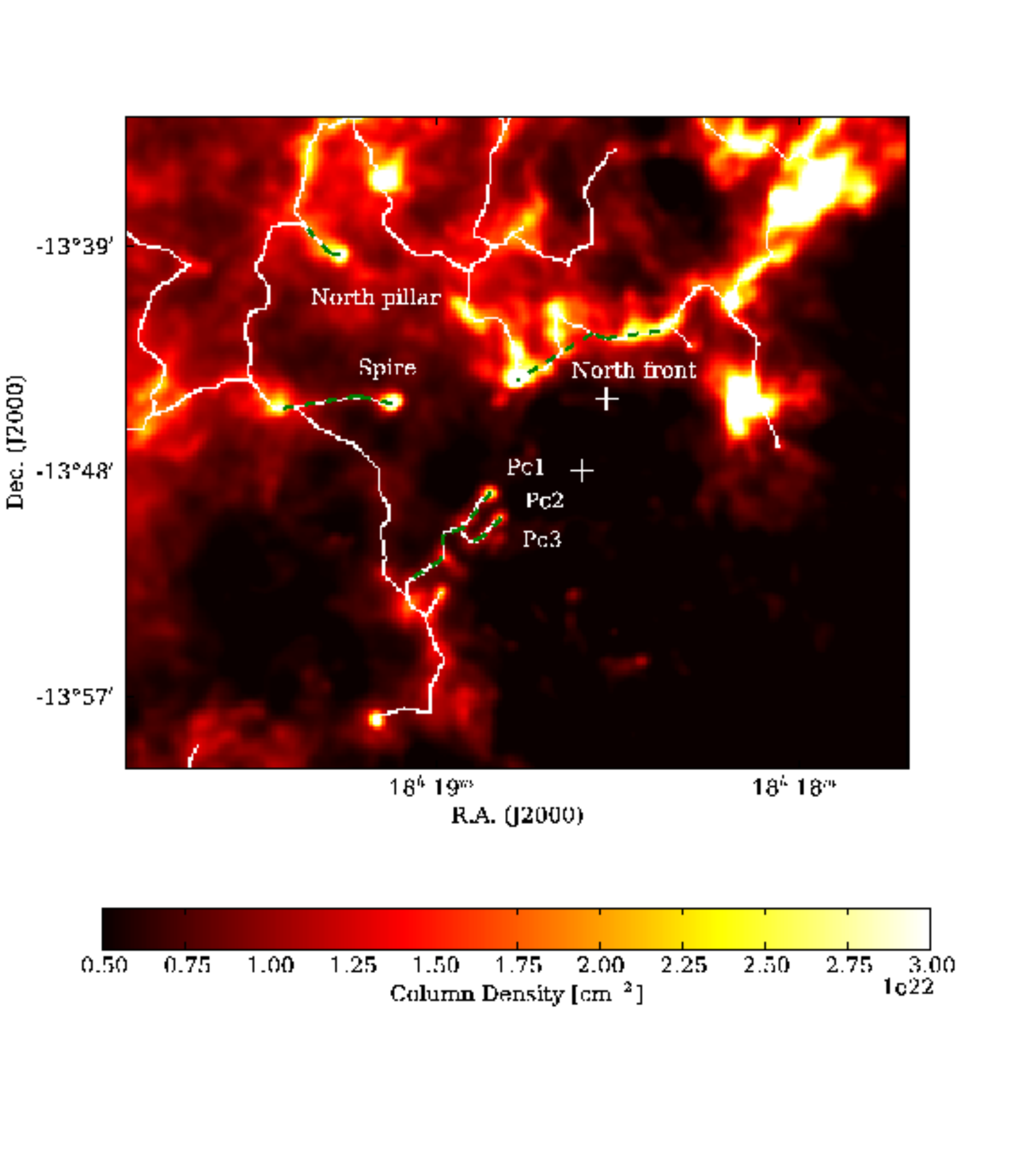}
\caption{\label{m16_N} Column density map of the interface between
  the Eagle Nebula and the \ion{H}{ii} region around NGC 6611
  (resolution of $25''$). The 
  DisPerSe skeleton is drawn in white. The intensity contrast level is 4$\times$10$^{21}$ cm$^{-2}$ and the
  low-density threshold is 5$\times$10$^{21}$ cm$^{-2}$ \citep[see][]{Sousbie:2011ft}, same
  values used for the Rosette map. The parts used for the
  profiles in Fig.~\ref{m16_shock_1} are indicated with the green
  dashed lines. The position of the main ionizing 
  sources (O4 and 05 stars) are indicated with the white crosses \citep{Oliveira:2008ve}.}
\end{figure}
\begin{figure}[t]
\centering
\includegraphics[trim=0 0 1.70cm 0 ,width=0.49\linewidth]{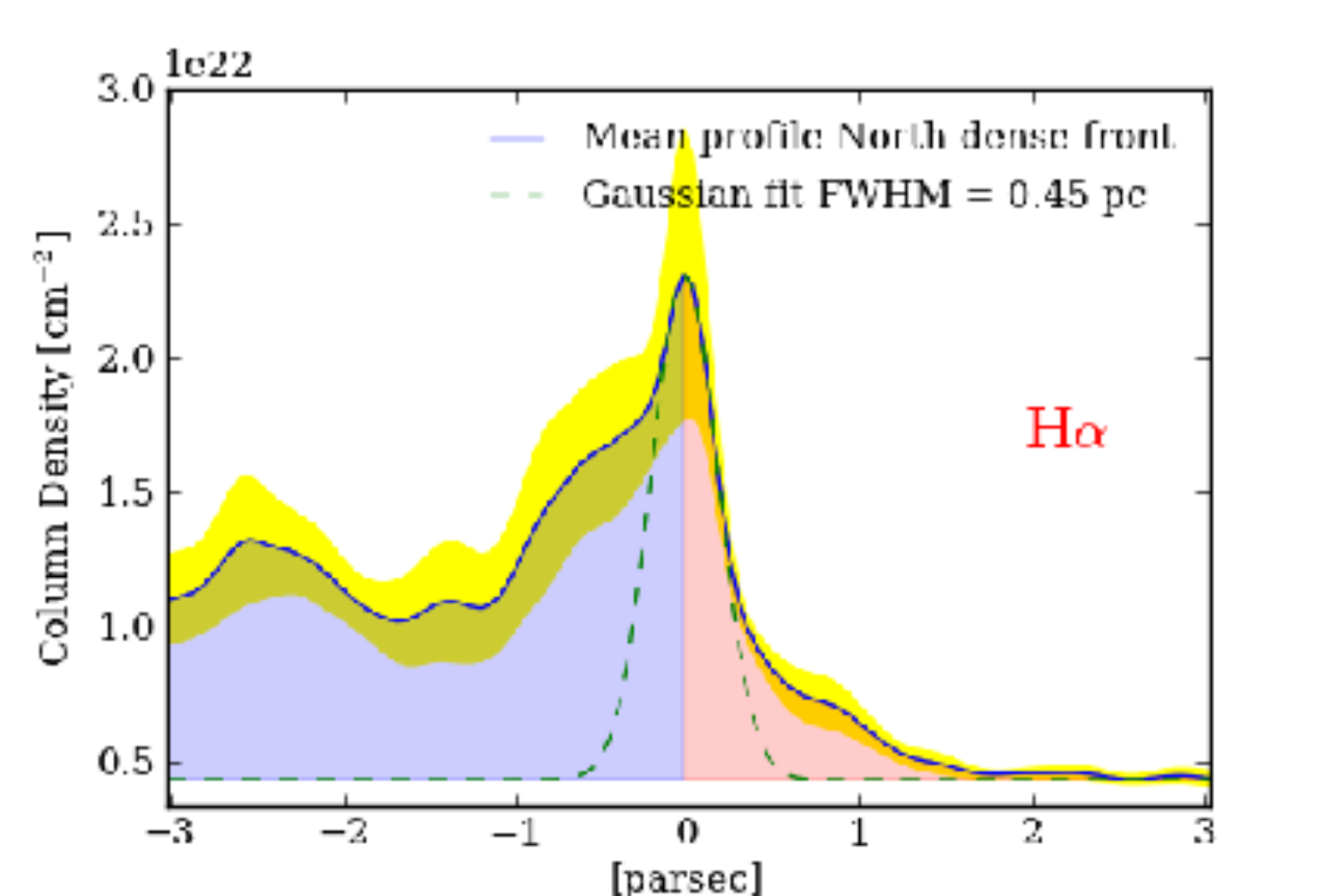}
\includegraphics[trim=0 0 1.70cm 0 ,width=0.49\linewidth]{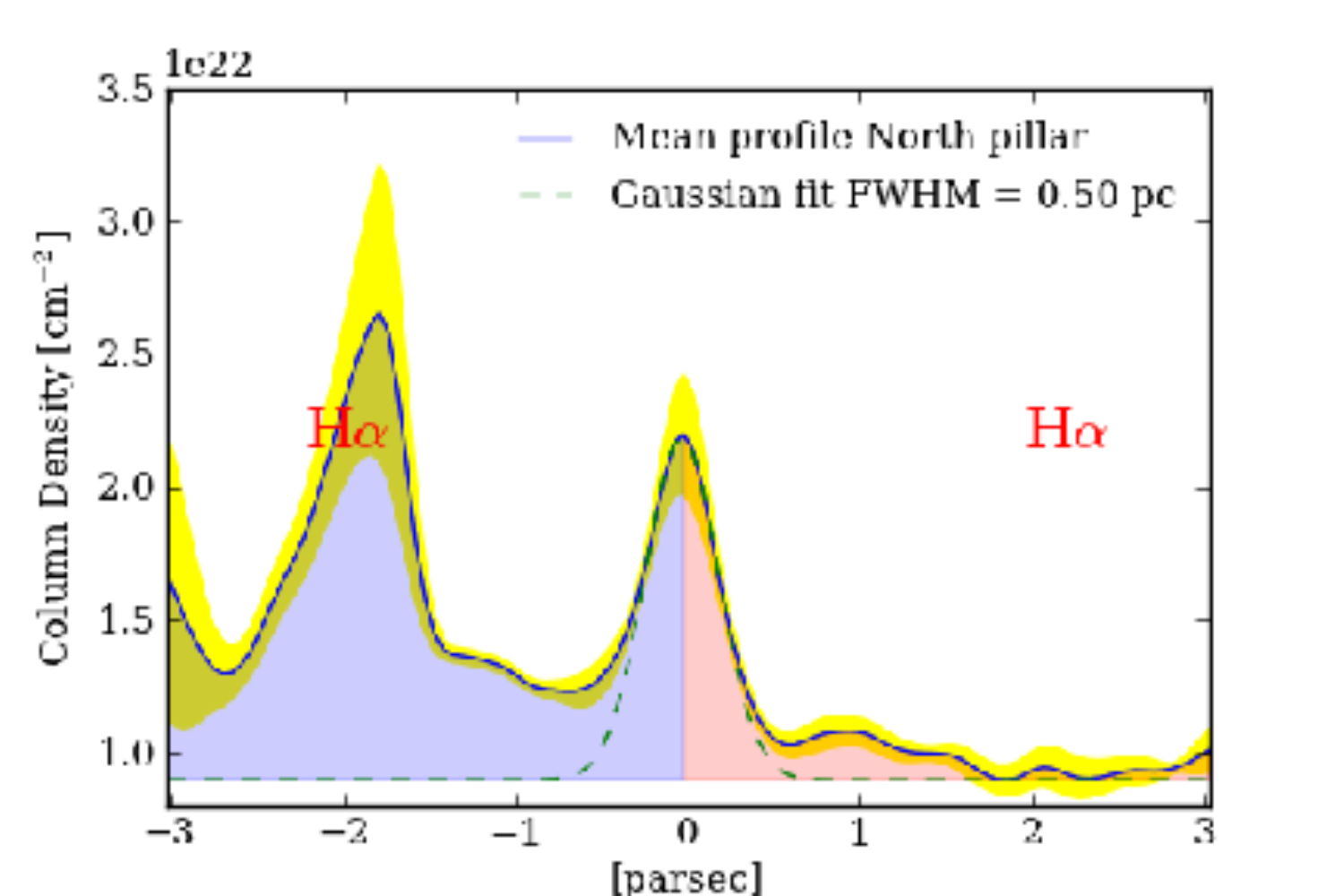}
\includegraphics[trim=0 0 1.70cm 0 ,width=0.49\linewidth]{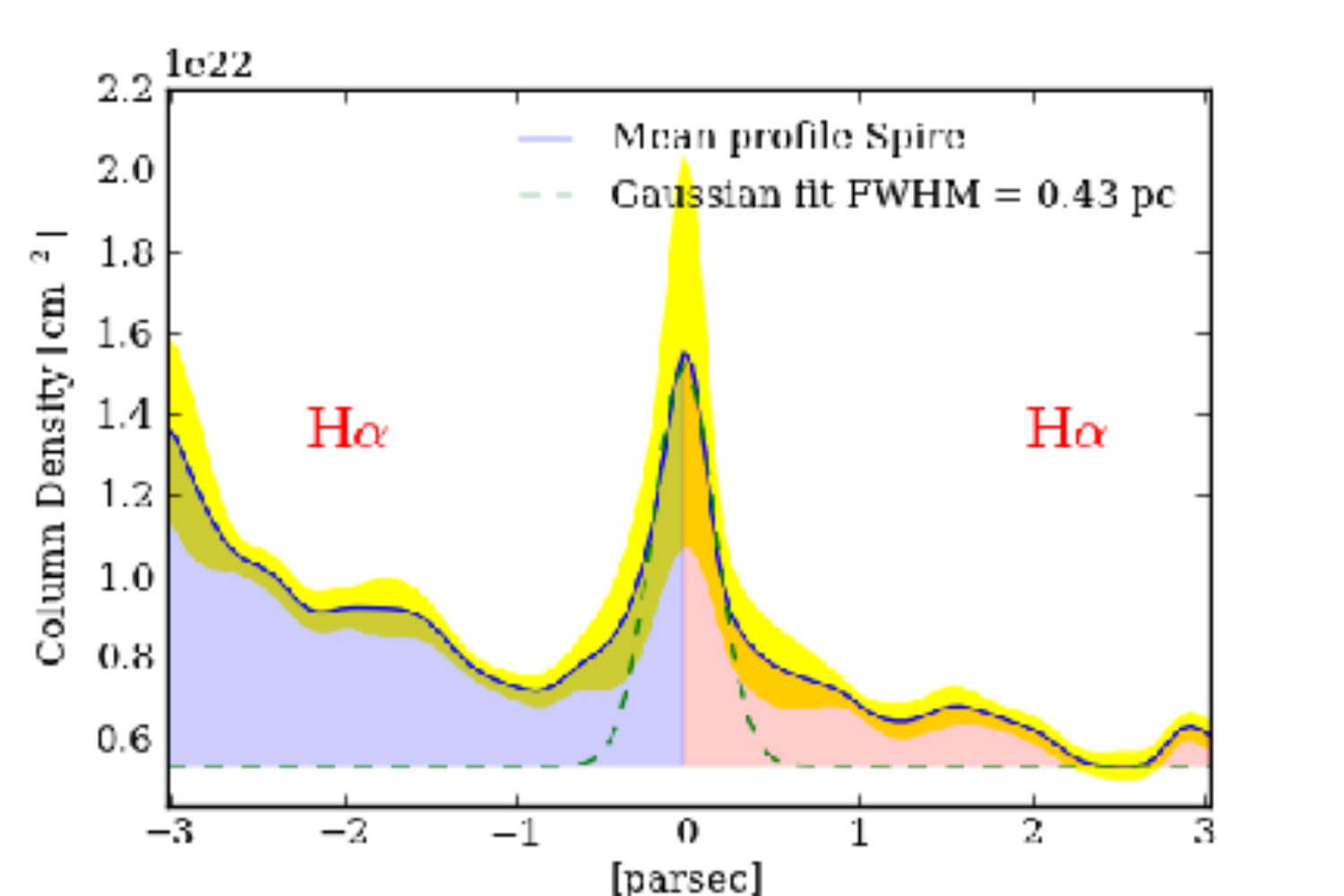}
\includegraphics[trim=0 0 1.70cm 0 ,width=0.49\linewidth]{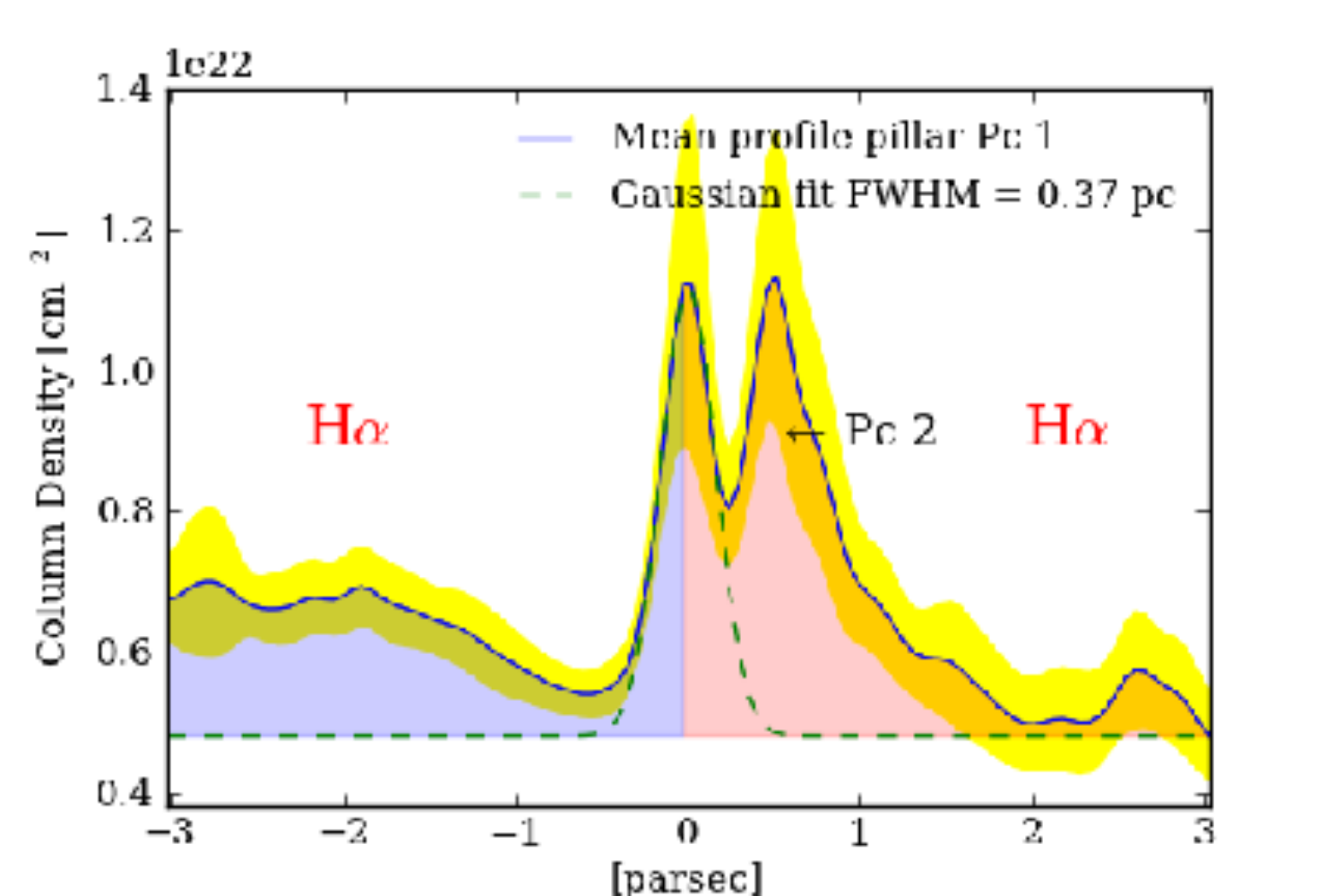}
\includegraphics[trim=0 0 1.70cm 0 ,width=0.49\linewidth]{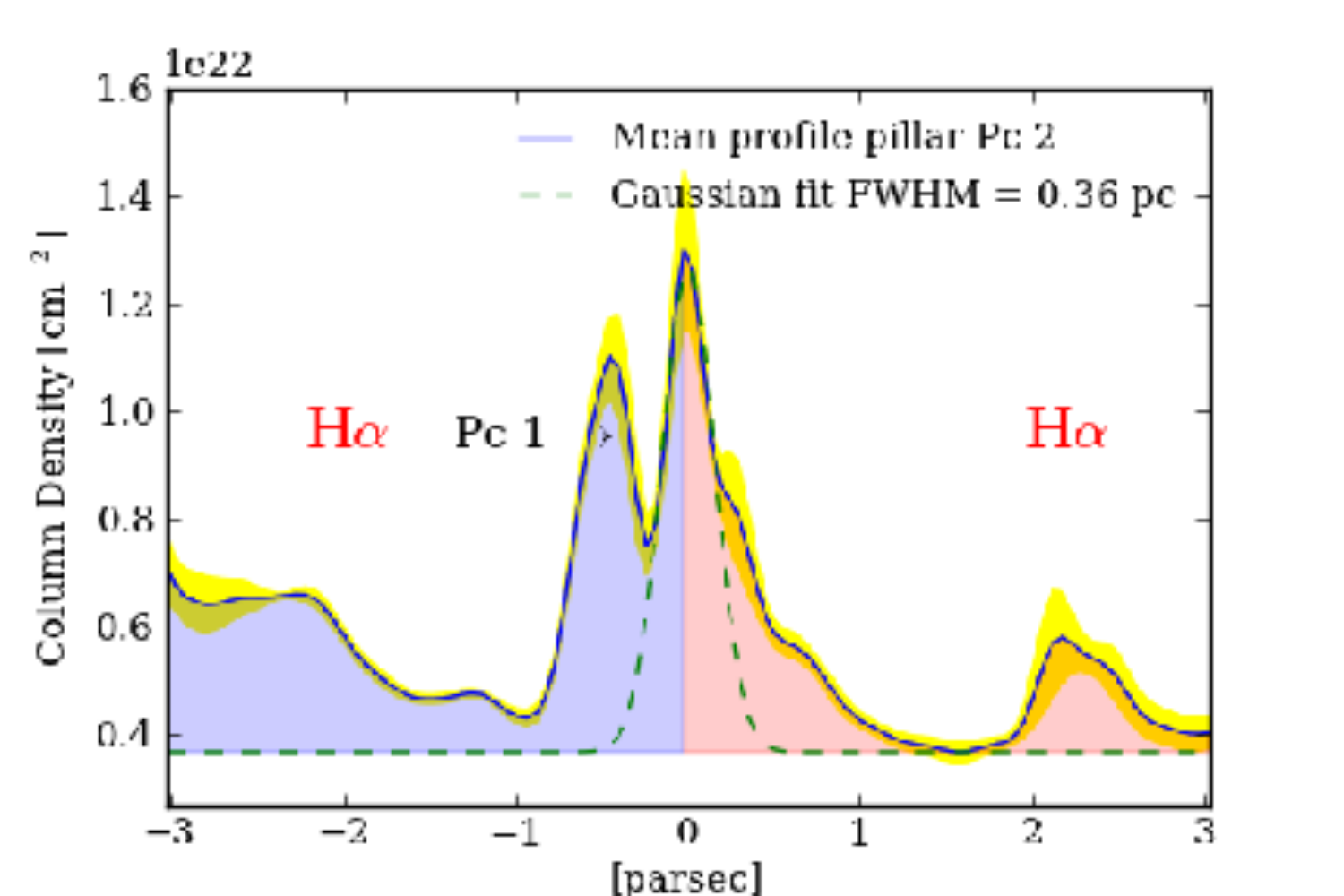}
\captionof{figure}{\label{m16_shock_1} Profiles of the dense front at the north
  of the main ionizing sources and profiles of the pillars traced by
  DisPerSe in M16. The
    extension of the profiles is fixed at 3 parsecs. The yellow shaded
    area represents the standard 
  deviation of the profiles. The red sign H$\alpha$ indicates on which side
  ionized gas is present (see also Fig.~\ref{m16_rgb_spectra} and the
  H$\alpha$ emission). The green-dashed curve is the Gaussian fit used
  to determine the FWHM width.}
\textcolor{white}{.}\\
\centering
\begin{tabular}{l|cccc}
dense front & FWHM (pc) & $\sigma_{FWHM}$ (pc) & Asymmetry & $\sigma_{asym}$ \\
\hline
\hline
North front  & 0.44 & 0.12 & 3.3 & 1.4 \\
North pillar & 0.54 & 0.08 & 3.4 & 1.1 \\
Spire        & 0.45 & 0.13 & 2.4 & 0.8 \\
Pc 1         & 0.38$^\star$ & 0.09 & 0.4 & 0.4 \\
Pc 2         & 0.35$^\star$ & 0.08 & 1.8 & 0.4 \\
\end{tabular}
\captionof{table}{\label{m16_param} Deconvolved FWHM width and asymmetric parameter for
  the averaged profiles of the 5 dense fronts of M16. $^\star$The width of the two
  largest pillars in the three Pillars of Creation is affected by a
  lack of angular resolution.}
\end{figure}

\section{Eagle nebula} \label{m16}

The Eagle nebula (M16) is located in the constellation of Serpens at
1.8 kpc from the Sun \citep{Bonatto:2006kb}. A young stellar cluster
NGC 6611 is ionizing the molecular gas of this star-forming region.
The principal ionizing sources are O4 and O5 stars whose combined
ionizing flux is of the order of $2\times 10^{50}$ s$^{-1}$
\citep{Hester:1996ir,White:1999ue}. The discovery of pillars of gas by
the {\it Hubble} space telescope \citep{Hester:1996ir} popularized
this region with the name of ``Pillars of Creation''. The region is
chemically rich, and spectral
  line studies of the region are numerous \citep[see][among
  others]{Pound:1998hj,White:1999ue,Allen:1999cl,Urquhart:2003jj,Schuller:2006ku,Linsky:2007bj}.
Recently, \citet{Flagey:2011jr} have studied the dust SED using {\it Spitzer} data.  They show that the
SED cannot be accounted for by interstellar dust heating by UV radiation,
but an additional source of heating is needed to match the mid
IR flux. They propose two possible origins:
stellar winds or a supernova remnant. {\it Herschel} provides new
observations of the region, and \citet{Hill:2012jb} studied
the impact of the ionizing source in detail on the initial temperature of pre-stellar
cores located at the edge or deep into the molecular
gas and did not find evidence of any extra source of
heating.

Even if the results of \citet{Flagey:2011jr} are confirmed in
the future, we do not expect that it will affect our
conclusions too much: we concentrate on the kinematic of the cold molecular gas
that should be relatively similar regardless of the dominant source of
energy (ionization thermal pressure, wind ram pressure, supernova
thermal pressure, etc.). The 70 $\mu$m 
  data are not used to derive the H$_2$ column density since it
  traces small grains in the PDRs instead of the cold dust \citep{Hill:2012jb}.
Figure~\ref{regions_over} shows as an 
  overview the SPIRE 250 $\mu$m map of the M16 cloud, reduced with
  the pipeline in HIPE10 (see Sect.~\ref{rosette}).  In Sect.
\ref{M16_cn} we investigate the column density structure of the edge
of the molecular cloud, using column density maps at a resolution of
$25''$ by ignoring the 500 $\mu$m emission. This choice is motivated
by the width of the pillars in the 70 $\mu$m, which is is smaller than their
width in the {\it Herschel} maps at an angular resolution of $37''$.
In Sect. \ref{M16_v} we concentrate on
the velocity structure of the head and the base of the three pillars
of  creation.

\subsection{Column density structure}\label{M16_cn}

 In \citet{Hill:2012jb}, the width
of the pillars of creation is not resolved in the
{\it Herschel} column density maps that have a resolution of $37''$
with the 500 $\mu$m data. Indeed, their width traced by the PDR in the
70 $\mu$m map is of the order of $25''$. By removing the longest
wavelength in the column density map, the
resolution can be improved to $25''$ (limited by the 350 $\mu$m data).
Figure~\ref{m16_N} shows the $25''$ {\it Herschel} column density map of the
dense cold regions at the edge of the \ion{H}{ii} region around NGC
6611. At this resolution, the two largest pillars of the pillars of
creation (Pc1 and Pc2), the Spire, and the pillar in the north of the
cavity are detected with the DisPerSe algorithm.  They appear as
disconnected filaments pointing towards the ionizing sources and
connected at the base to the rest of the molecular cloud by a triple
point in the skeleton. The DisPerSe algorithm connects the
  largest pillar, Pc1, to the rest of
the molecular cloud, although the connection is not
 obvious in molecular line observations
\citep[see][]{White:1999ue}. However, the velocity also suggests that
the pillar is still or was
connected to the rest of the cloud (see Sect. \ref{M16_v}). The
third pillar (Pc3) of the three
pillars of creation is not detected, probably because of the resolution or the
threshold in the DisPerSe algorithm. A
nascent pillar is also detected in the south of the pillars of creation. 

Figure~\ref{m16_shock_1} presents the averaged column density profiles
of the pillars and the profile of the northern front. In the northern
region, the compression is clearly visible thanks to the asymmetry of the
profile (more than a factor of three between the two sides of the
crest, see Table~\ref{m16_param}).
The profiles of the pillars are also asymmetric; however, there is no
preferred side, because it depends on the confusion with other structures in
the profiles (e.g. Pc1 and Pc2 have reversed asymmetries because they
both contribute to the profile of each other). The deconvolved FWHM is also
indicated in Table~\ref{m16_param}. The widths of the two largest
pillars of creation are $44''$ and $40''$, which is almost twice their size as
measured in the {\it Herschel} 70-$\mu$m map (resolution of 
$\sim$6$''$). The width of the pillars is not properly resolved even
if the measured size is twice the size of the beam at 350 $\mu$m
($25''$ resolution).

\begin{figure}[t]
\centering
\includegraphics[trim=0cm 1cm 2cm 0cm,width=\linewidth]{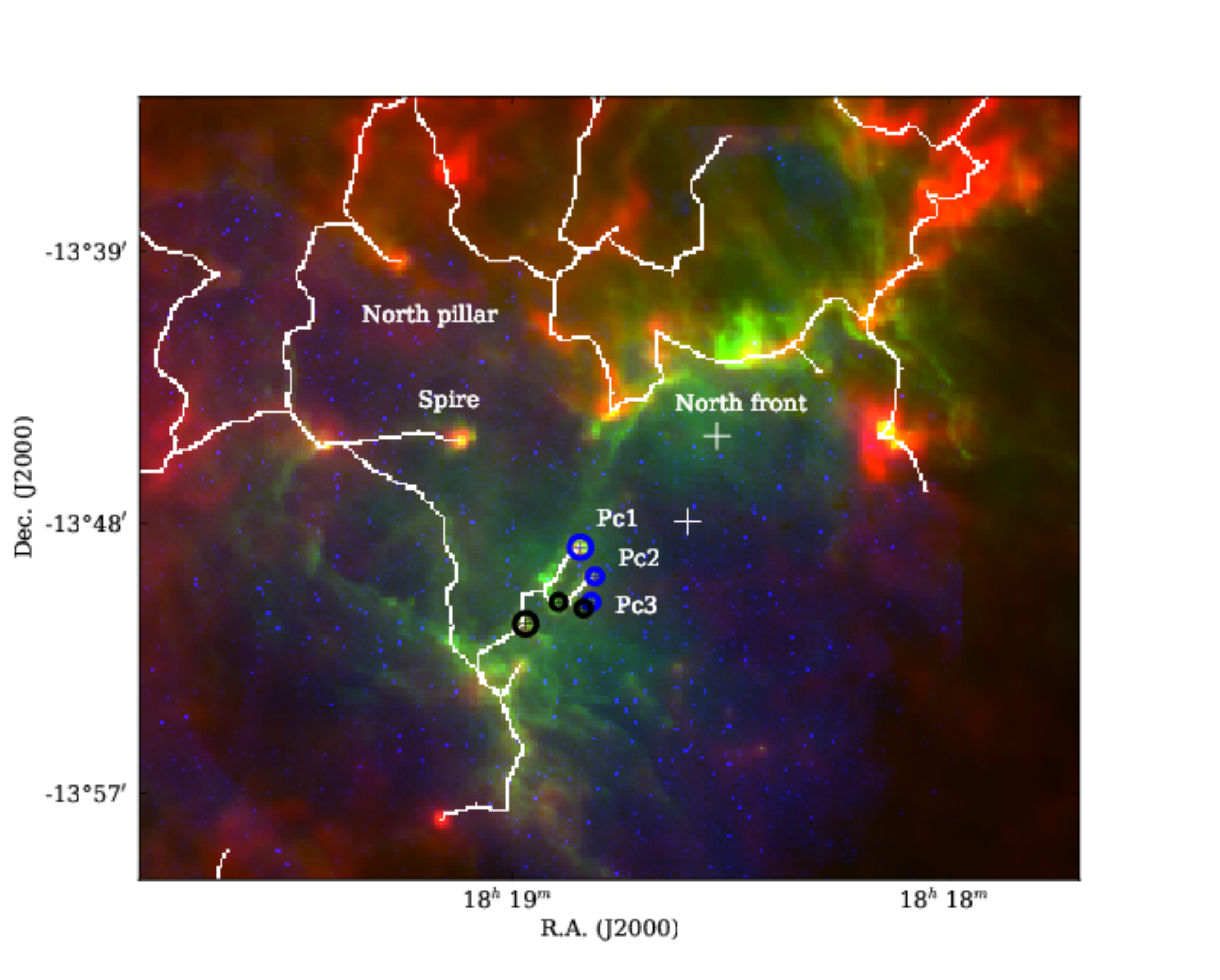}
\caption{\label{m16_rgb_spectra} Three-color image of the interface between
  the Eagle Nebula and the \ion{H}{ii} region around NGC 6611 (red:
  {\it Herschel} column density map, green: PACS 70 $\mu$m, blue: H$\alpha$). The
  white crosses indicate the O stars. Each spectrum in
  Fig.~\ref{m16_velocity} is integrated inside the corresponding
  colored circle.}
\end{figure}

\begin{figure}[t]
\centering
\includegraphics[trim=0 0 0cm 0
  ,width=0.8\linewidth]{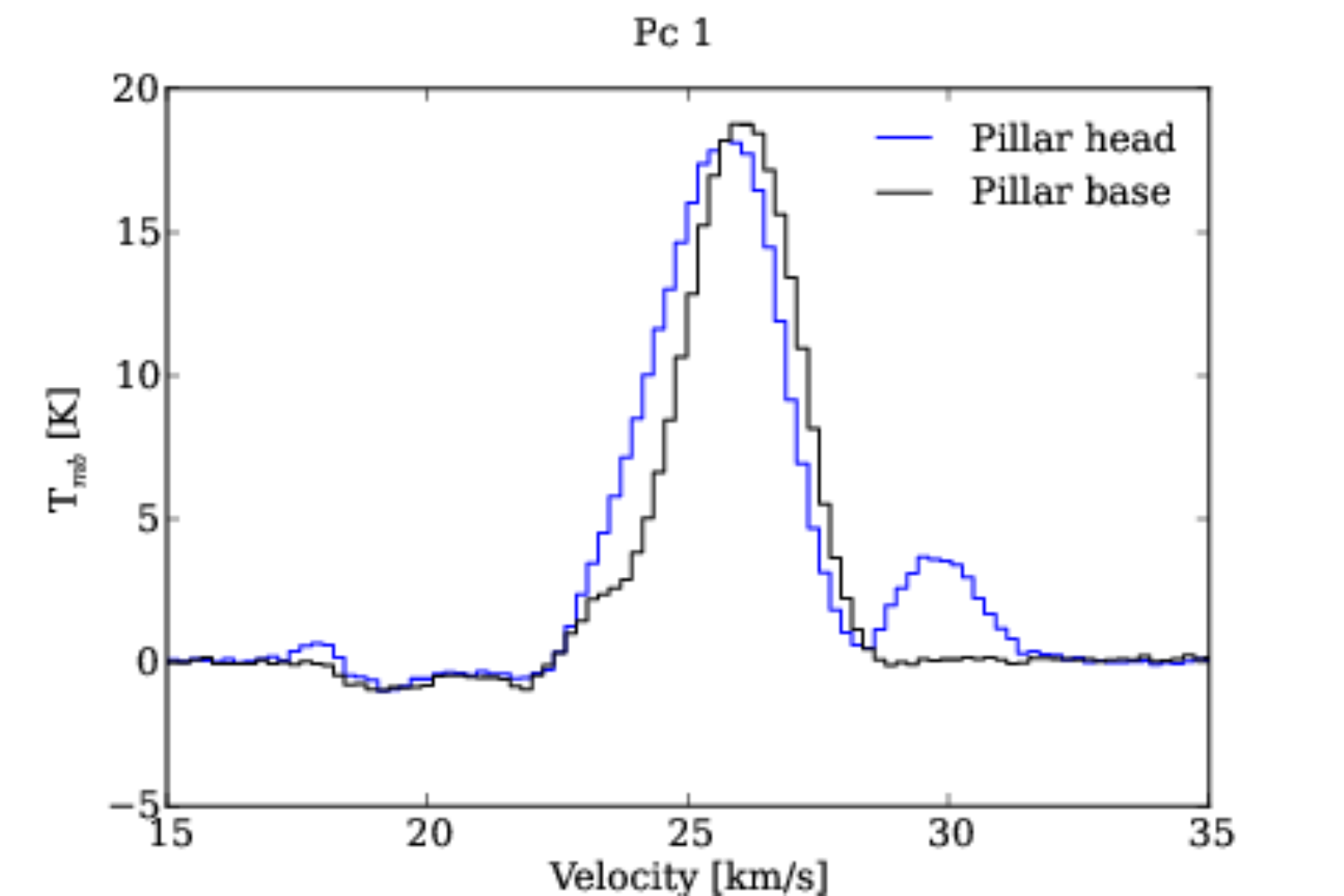}
\includegraphics[trim=0 0 0cm 0
  ,width=0.8\linewidth]{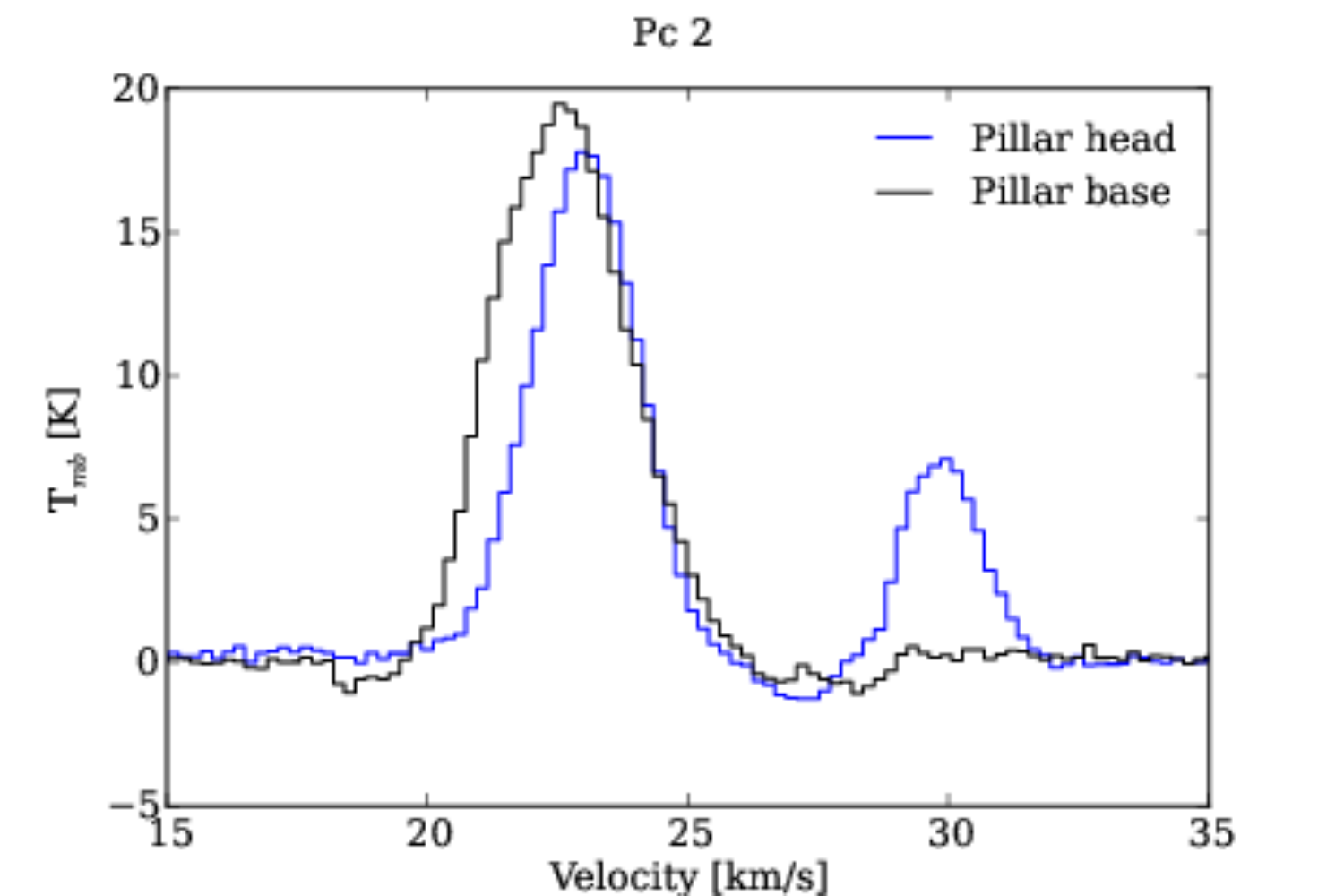}
\includegraphics[trim=0 0 0cm 0
  ,width=0.8\linewidth]{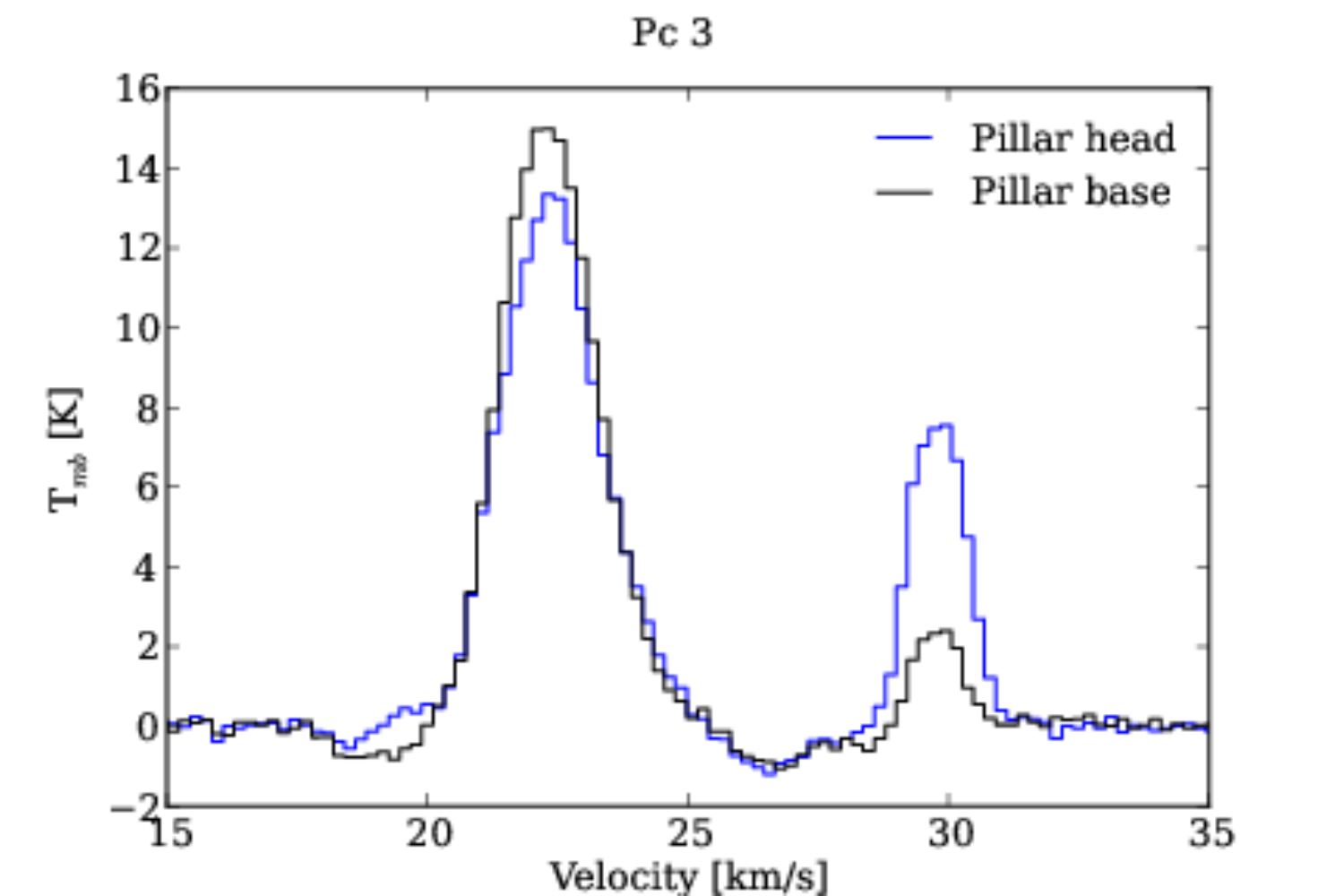}
\caption{\label{m16_velocity} $^{12}$CO 3-2 integrated
spectra on the three Pillars of Creation. The spectra are
spatially integrated in the colored circles showed in
Fig.~\ref{m16_rgb_spectra}.}
\end{figure}

\subsection{Velocity structure}\label{M16_v}

There are only a few high angular resolution spectral-line observations of the whole M16 nebula.
The majority of the studies concentrate on the three pillars of creation and SFO30
\citep[the north of the cavity, see][]{Oliveira:2008ve}. Using data
from the James Clerk Maxwell Telescope (JCMT\footnote{JCMT is operated
  by the Joint Astronomy Centre on behalf of the Science and
  Technology Facilities Council of the United Kingdom, the Netherlands
  Organization for Scientific Research, and the National Research
  Council of Canada.}) public archives, we examine the velocity
structure of the pillars of creation with $^{12}$CO
(3-2). Figure~\ref{m16_rgb_spectra}  is a three-colour (Red: {\it Herschel} column density
- Green: 70 $\mu$m {\it Herschel} - Blue: H$\alpha$) image around the ionized
bubble, where we have indicated the different areas that we
selected based on the availability of 
$^{12}$CO (3-2) spectral-line data taken at JCMT. In this work we have
analyzed each pillar individually and looked at the differences in the
velocity profiles between the head and the base of each of them (see
Fig.~\ref{m16_velocity}). The blue spectra are integrated around the
head of the pillars, and the black ones are integrated around the base.

Pc1 has a bulk velocity around +26 km s$^{-1}$. As previously mentioned, most observations of the pillars of creation display a gap between the
three heads and the rest of the cavity. (The gap is just to the north
of the black circle at the base of Pc1 in Fig.~\ref{m16_rgb_spectra}.)
The DisPerSe algorithm, however, connects the pillars to the edge of the cavity
based on the contrast level that was used. The connection along Pc1
is also supported here by the velocity spectra that look alike between
the two sides of the ``gap'', while Pc2 and Pc3 have a different bulk
velocity around +22.5 km s$^{-1}$. The difference oinbulk velocity
between Pc1 and Pc2/3 has already been seen by \citet{White:1999ue} and may
be interpreted as a difference in the orientation and location of the
pillars. Indeed, in the {\it Hubble} image of \citet{Hester:1996ir},
Pc1 appears illuminated, while the Pc2 and Pc3 appear dark. This
suggests that 
Pc1 is located behind the ionizing source at the back of the cavity,
while Pc2 and Pc3 are between the ionizing source and the observer.
The three spectra that are integrated around the heads of the pillars
also show a velocity component at 30 km s$^{-1}$, which probably
corresponds to a line-of-sight cloud as claimed by \citet{White:1999ue}.

The two longest pillars Pc1 and Pc2 have small velocity gradients between the head
and the base (see Fig.~\ref{m16_velocity}). For Pc1, the head of the
pillar is red-shifted at + 0.4 km~s$^{-1}$ relative to the base, and for Pc2,
the base is red-shifted at + 0.4 km~s$^{-1}$ relative to the
head. 
These gradients are projections on the line of sight and could
be the result of a change in orientation of the velocity between the
head and the base. However, such a gradient perpendicular to the axis
of the pillar would mean that the head covered a distance of 0.2 pc in
500 ky away from the rest of the pillar, leading to its destruction. Considering the age
of the central cluster, it is 
much more probable that the velocity field has a constant orientation along
the pillar, and the gradients can be interpreted as an increase or
decrease in the velocity along the pillar \citep[see also][]{Pound:1998hj}.  
 Considering the H$\alpha$ illumination, the head of Pc1 is
between its base and the observer, therefore the gradient can be
interpreted as an extension and a growth of the pillar. Since for Pc2,
both the orientation and the gradient are reversed, the pillar also
appears to be growing in length. No gradient is observed for Pc3, however the
pillar is small and the
resolution may not be sufficient to measure a possible gradient between
the head and the base of the pillar.

\section{Discussion and comparison with numerical simulations}\label{discussion}

\subsection{Large-scale compression from \ion{H}{ii} regions}

The dense fronts around the ionized gas in the Rosette
and Eagle nebula are systematically asymmetric with low-density gas
where H$\alpha$ emission is present.
This is an indication that these regions are compressed by the ionized
gas. This result is similar to those of \citet{Peretto:2012bu} for the Pipe nebula,
  \citet{Schneider:2013bu} for Orion B, and \citet{NguyenLuong:2013dp} for
W43-Main. They all found that the gas
is compressed by the influence of near-by OB associations.  In our two regions, the
low density corresponds to the ionized gas from which the
material has been swept out, the dense front accumulates the evacuated
gas in a dense shell, and the molecular cloud is unperturbed behind
the front.

\begin{figure}[t]
\centering
\includegraphics[trim=30 2 2cm 0cm
  ,width=\linewidth]{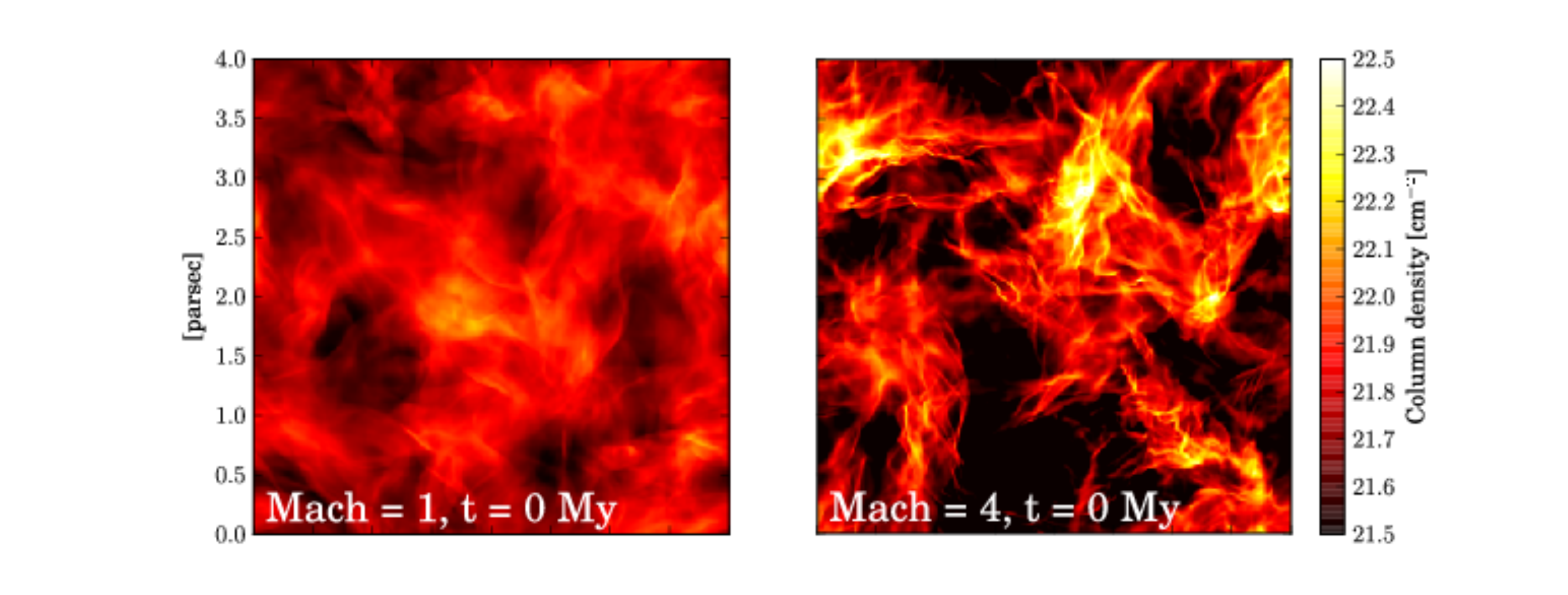}
\includegraphics[trim=30 2 2cm 0cm
  ,width=\linewidth]{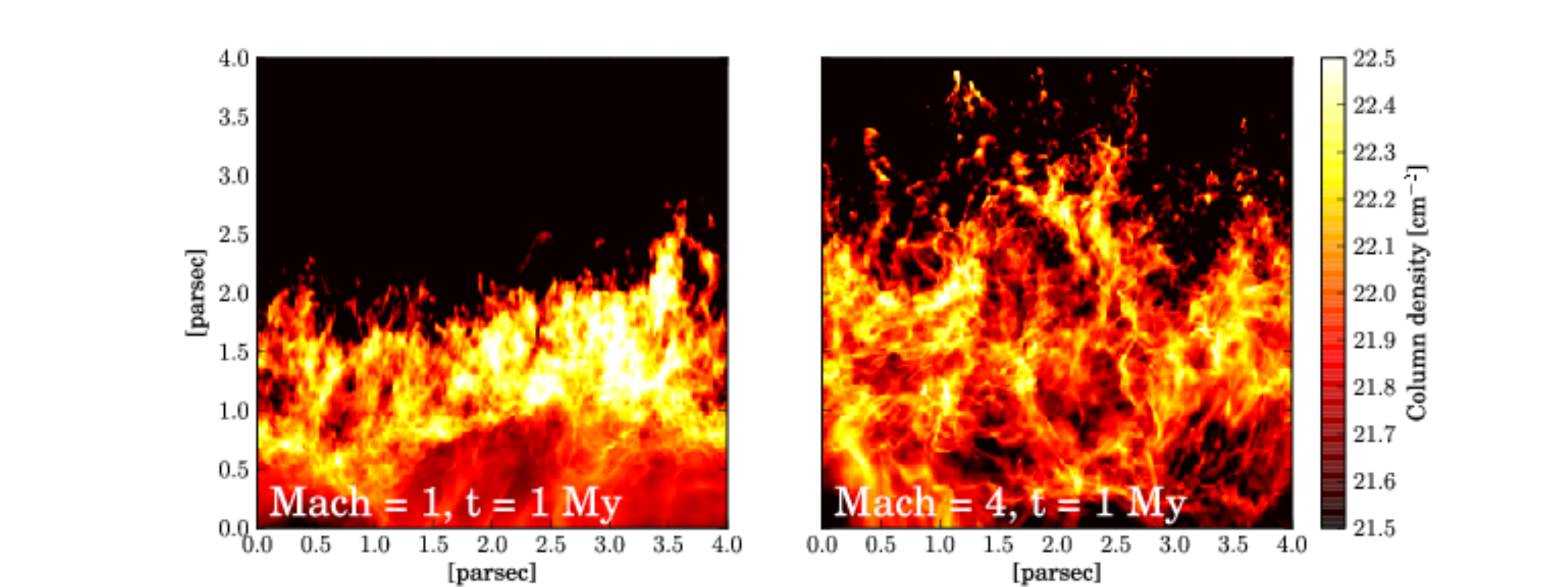}

\caption{\label{simu_turb_N} Column density snapshots of the
  ionization of a turbulent medium at
  Mach 1 and Mach 4
  \citep[see][]{Tremblin:2012he}.  The boxes are cubes of 4-pc side at
  a resolution of 0.01 pc per cell. The initial
  conditions are a mean density of 500
  cm$^{-3}$ and a ionizing flux of $10^{9}$~ Lyc photons s$^{-1}$~cm$^{-2}$ coming
  from the top of the boxes. Top:
  before the ionization is 
  introduced. Bottom: after 1 My.} 
\end{figure}

\begin{figure}[t]
\centering
\includegraphics[trim=0 0 0cm 0
  ,width=0.8\linewidth]{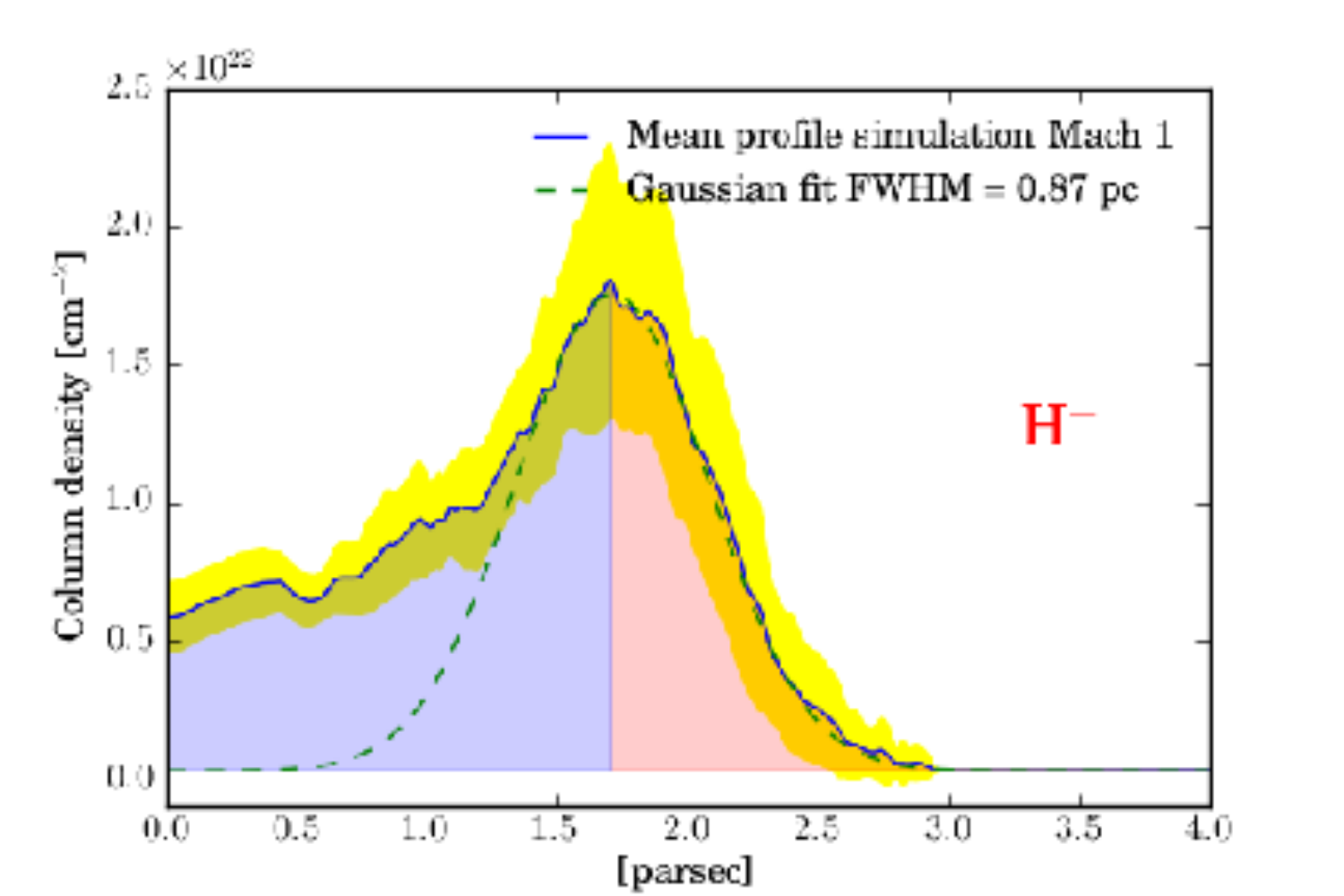}
\includegraphics[trim=0 0 0cm 0
  ,width=0.8\linewidth]{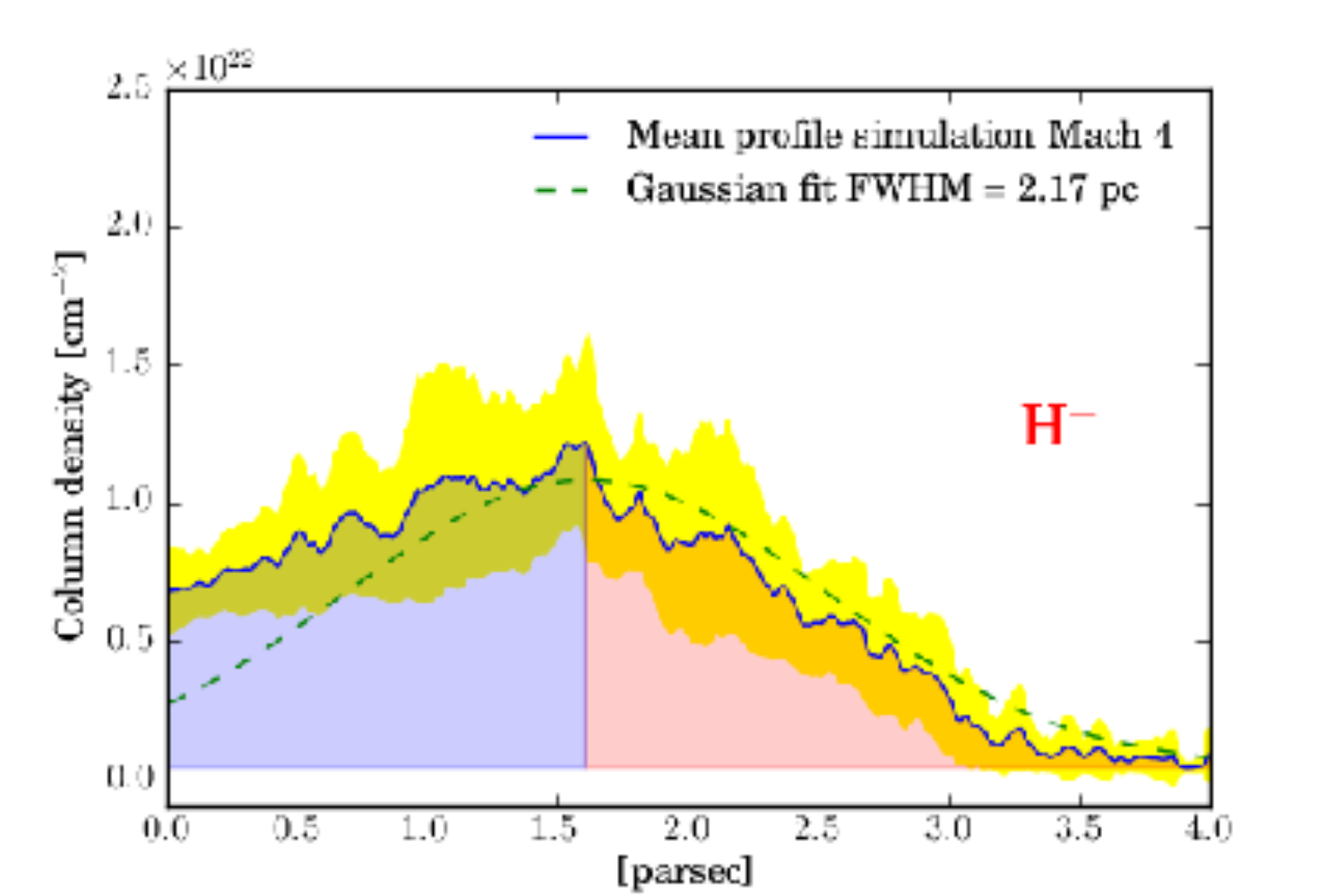}
\caption{\label{simu_asym} Column density profiles of the two
  simulations shown in fig. \ref{simu_turb_N} after 1 My. The red sign
H$^+$ indicates the position of the ionized gas in the simulation. The
flux is coming from the right side of the profiles. } 
\end{figure}

 Such compression is also seen in numerical simulations (see Fig.~\ref{simu_turb_N}). The
mean column density profile of the ionization of a turbulent medium with 
Mach numbers of 1 and 4 (RMS velocity of 0.6 and 2.4 km/s) and a mean density of 500 cm$^{-3}$ is shown in
Fig.~\ref{simu_asym}. The simulations include neither gravity
  nor magnetic fields,
  the resolution is 0.01 pc, and the heating and cooling function of
  the cold molecular gas is taken from \citet{Wolfire:1995gp}. The
  heating by the diffuse radiation field in the PDR is neglected, and the
  heating by the ionization and the cooling by 
  recombination are taken in account when the gas is ionized. 
The width of the profiles and the compression at
the peak position are set by the
ratio of the ionized pressure to the turbulent ram pressure
\citep[see][]{Tremblin:2012he}. In the Mach-1 simulation, the
ionized-gas pressure is greater than the turbulent ram pressure,
and the compression is efficient, whereas in the Mach-4 simulation,
they are within an order of magnitude, and there is almost no
compression. Therefore, even if these simulations are not performed in 
the exact same conditions as the Rosette and Eagle nebulae, it is
likely that the ionized-gas pressure is
dominant at the edge of these \ion{H}{ii} regions and leads to the
observed compression. Assuming a Larson relation $v_\mathrm{RMS} = 1.1 \mathrm{km/s}
  (L_\mathrm{box}/\mathrm{pc})^{0.4}$, the RMS velocity on the scale
  of the box should be around 2 km/s. However, the measured profiles
  seem to indicate that the level of turbulence is closer to an
  RMS velocity of $\approx$ 0.6 km/s, this could indicate that the
  Larson relation is not suitable inside a single molecular cloud, as
  already proposed by \citet{Lombardi:2010ew} for the Larson third law.
These compressed profiles are close to what could be expected from the
collect process in the collect and collapse scenario proposed by
\citet{Elmegreen:1977iq} and could lead to a sequential star
formation. The observed ratio of the column density of the 
dense fronts to the column density of the unperturbed gas is in the range 4-5 in both regions (background subtracted). If we assume the
same density ratios and also dthat the gas is isothermal, the shell velocity
given by the Rankine-Hugoniot shock-jump condition would be at $\approx$ 1 km/s
($v_\mathrm{shell}^2/c_0^2=n_\mathrm{shell}/n_0$, with $c_0$
  the sound speed in the cold gas). This is too low for the Rosette nebula whose shell velocity was
estimated by \citet{Kuchar:1993bn} at 4.5 km~s$^{-1}$. However, we have seen in
\citet{Tremblin:2012he} that the density compression is in the range 10-100 in the Mach-1 simulation, whereas the column-density compression is
affected by the projection effect and is only about 3-4 (see
Fig.~\ref{simu_asym}). Therefore it is likely that the
column-density compression in the Rosette and M16 molecular clouds is
also affected by the projection effect and cannot be used to infer the
local density compression. 

Another indication of the collect process is the difference in the average
thickness between the dense fronts 1, 2, and 4 ($\approx$ 0.75 pc) and the
dense front 3 and the pillar in the Rosette nebula ($\approx$ 0.55 pc). The
standard deviation of the thickness along the fronts 
is quite large for the dense front 3, therefore this result needs
further investigation.The 
fronts 1, 2, and 4 are perpendicular to the propagation of the shell and therefore
can collect a large amount of material on the way along their length.
For the pillar and the front 3, they are parallel to
the direction of expansion and can only collect material along their
width, which is relatively small. Therefore the mass of these structures
cannot grow as much as the rest of the shell. This could
explain why their width is smaller than the width of the other
dense fronts that should be able to grow as a function of
time. 

\subsection{From nascent to old pillars}

\begin{figure}[t]
\centering
$\vcenter{\hbox{\includegraphics[trim=9cm 0 0cm 0cm
  ,clip,width=0.45\linewidth]{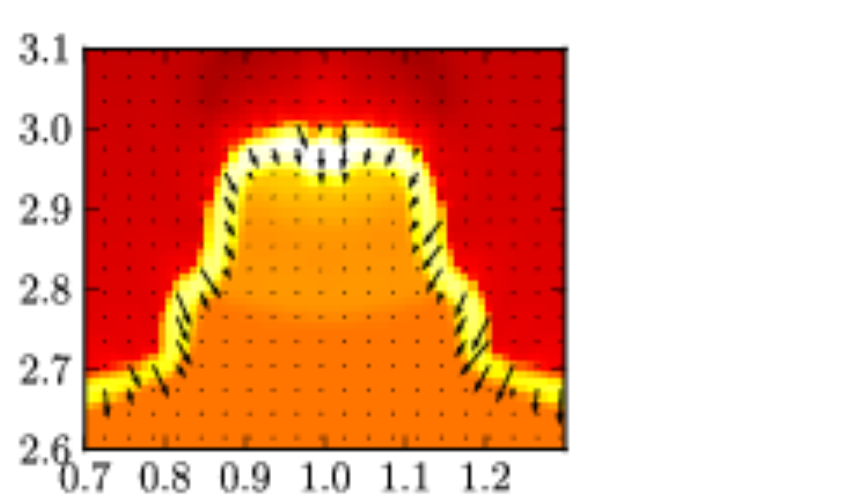}}}$
$\vcenter{\hbox{\includegraphics[trim=0 0 0cm 0
  ,width=0.45\linewidth]{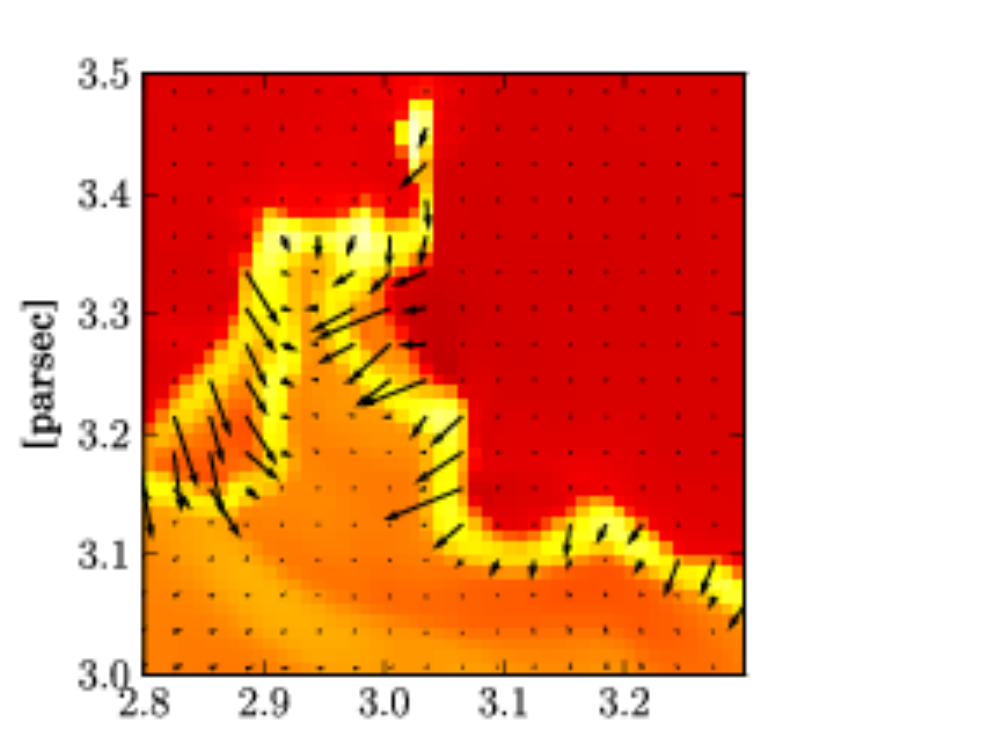}}}$
$\vcenter{\hbox{\includegraphics[trim=9cm 0 0cm 0
  ,clip,width=0.45\linewidth]{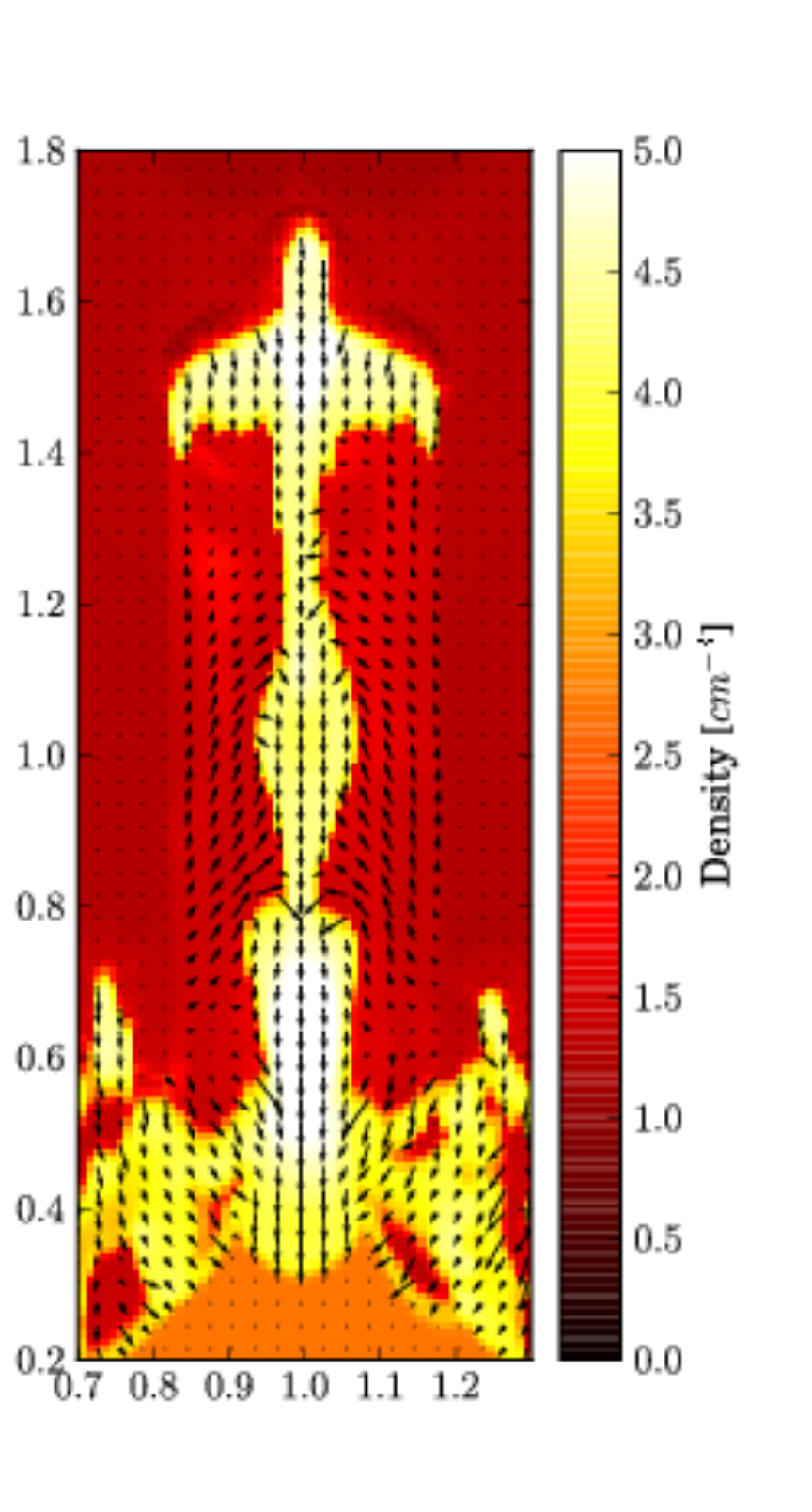}}}$
$\vcenter{\hbox{\includegraphics[trim=0 0 0cm 0cm
  ,width=0.45\linewidth]{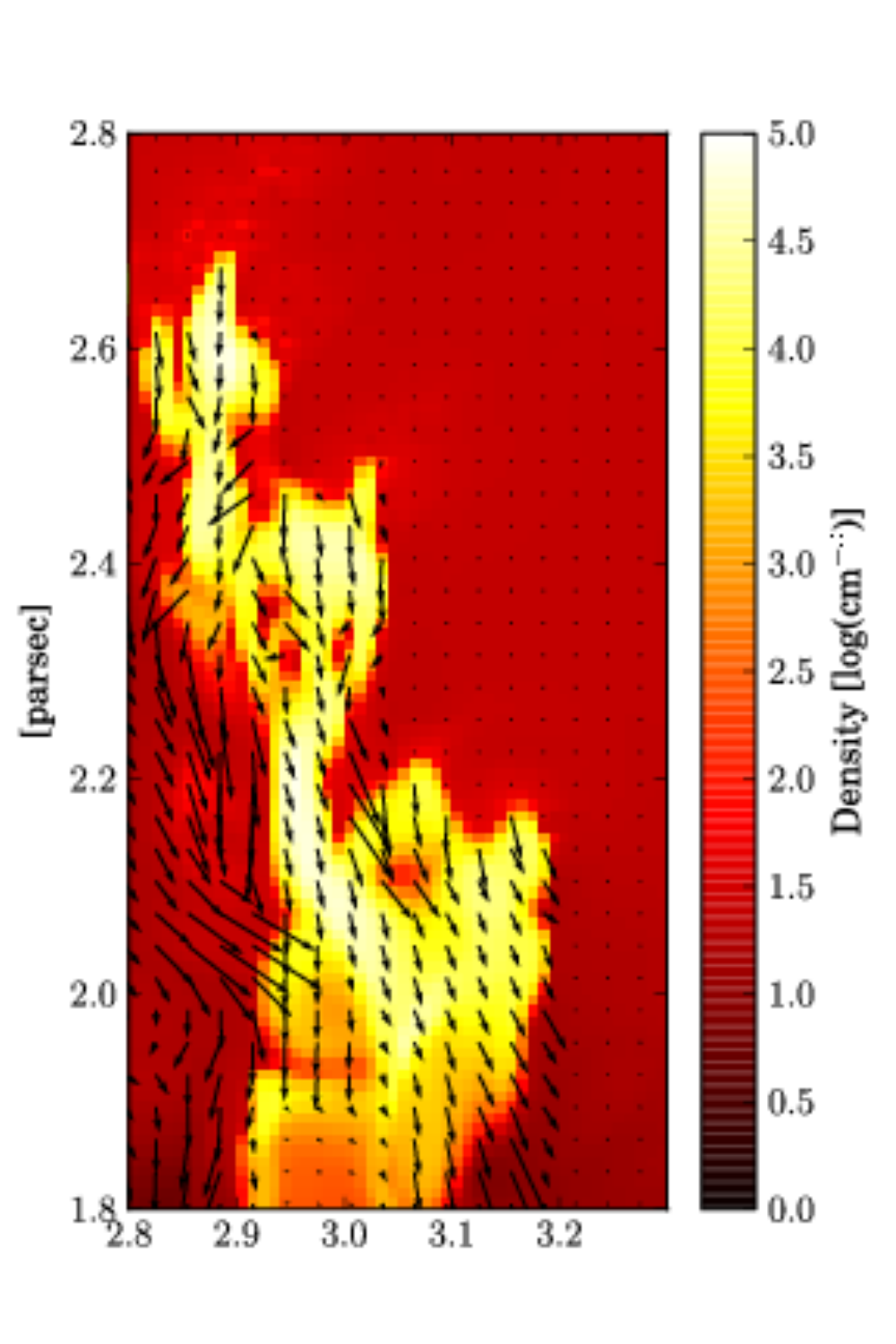}}}$

\caption{\label{simu_pillar_N} Density cut of the formation of a
  pillar in an idealized simulation 
  of a clump embedded in a homogeneous medium
  \citep[left :][]{Tremblin:2012ej} and in a turbulent medium at
  Mach 1 \citep[right:][]{Tremblin:2012he}. The black arrows
  represent the velocity field of the neutral gas. The compressed
  layer can be identified with the velocity field and the high density
  (around $10^4$ cm$^{-3}$ in yellow). Top: the snapshots are
  taken when the pillar is forming, just before the compressed layer has
  collapsed in on itself. The high curvature of the compressed layer can be
  clearly identified in both cases. Bottom: after the shell has collapsed, the
  pillars are formed.} 
\end{figure}

\begin{figure}[t]
\centering
\includegraphics[trim=0 0 0cm 0
  ,width=0.49\linewidth]{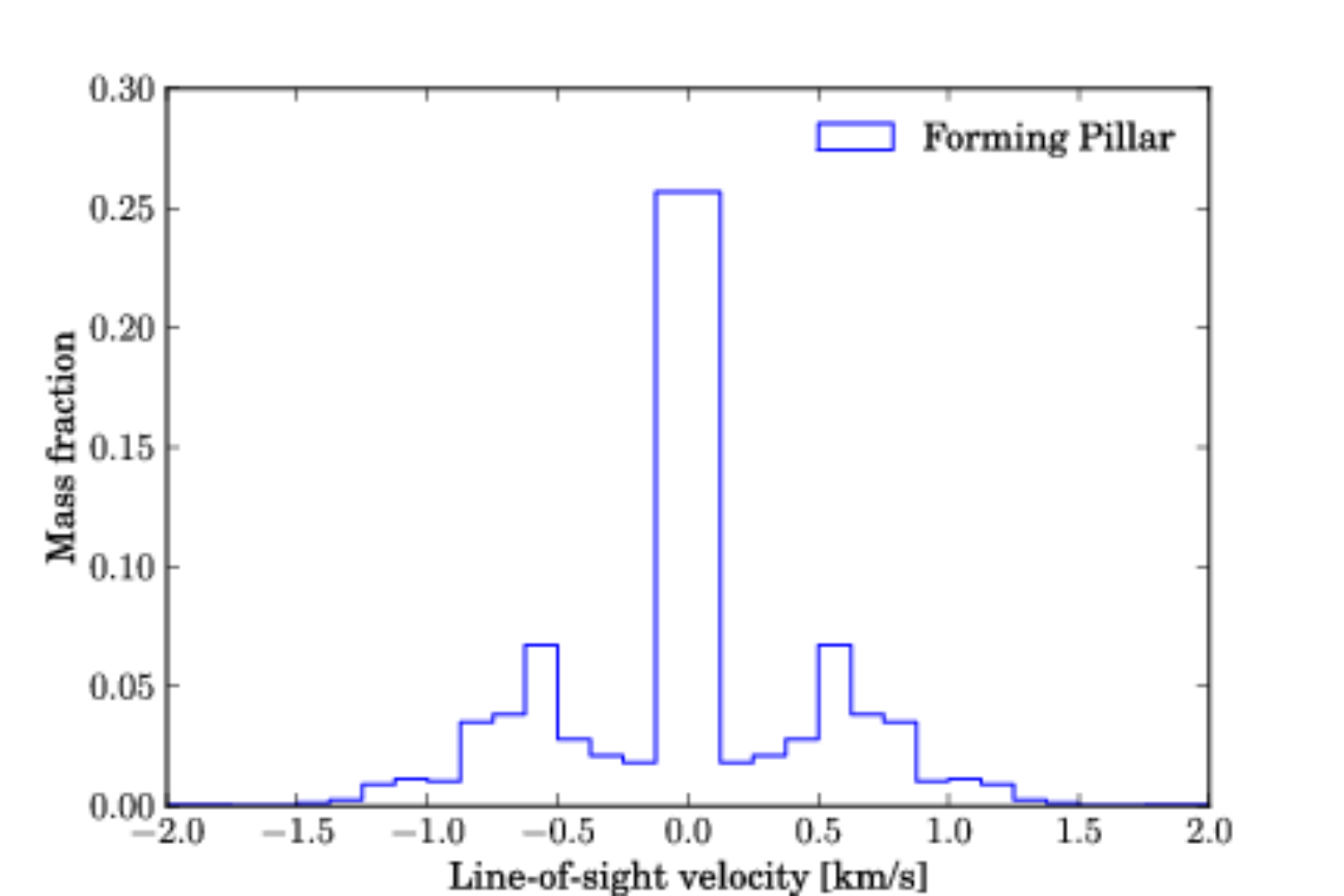}
\includegraphics[trim=0 0 0cm 0
  ,width=0.49\linewidth]{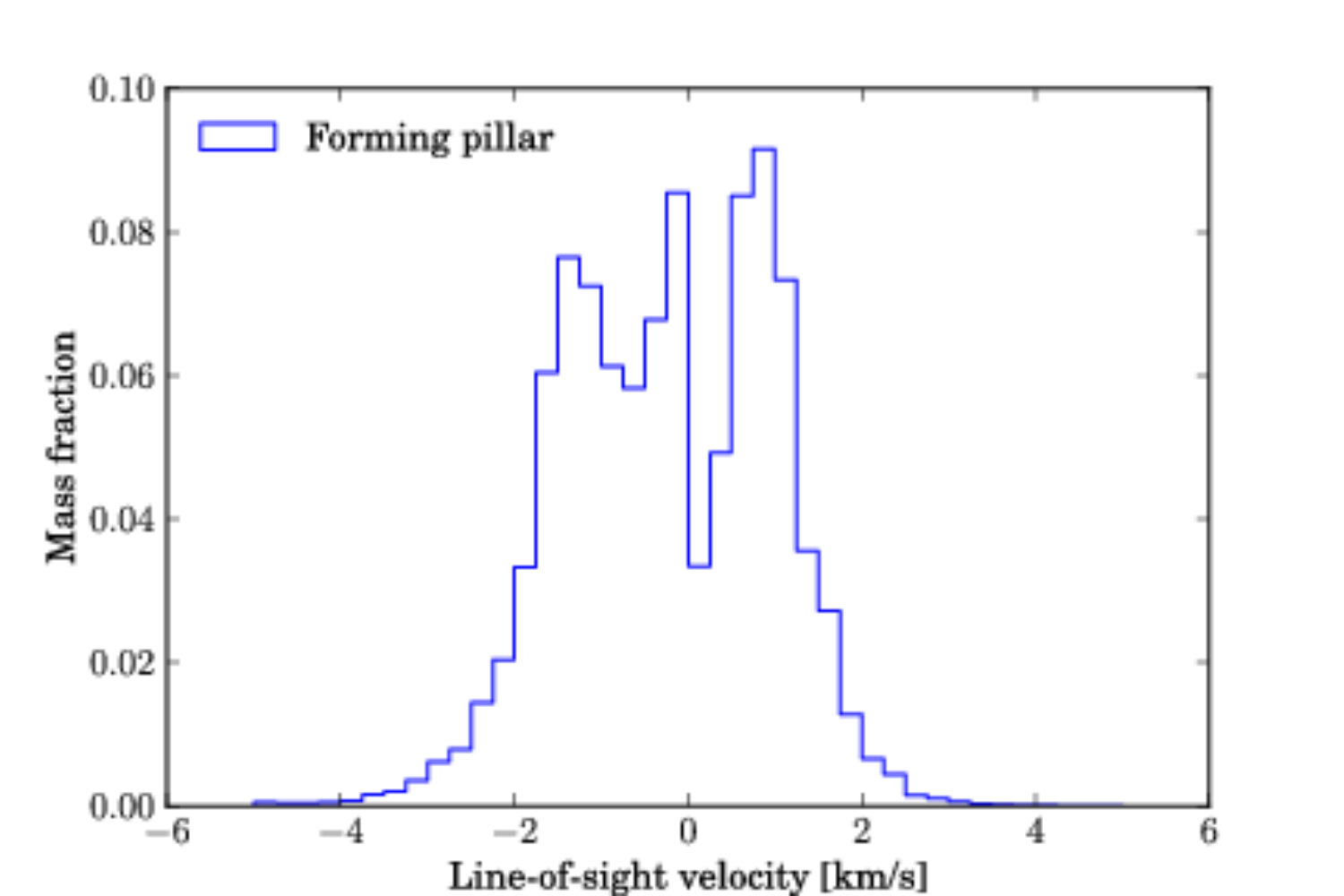}
\includegraphics[trim=0 0 0cm 0
  ,width=0.49\linewidth]{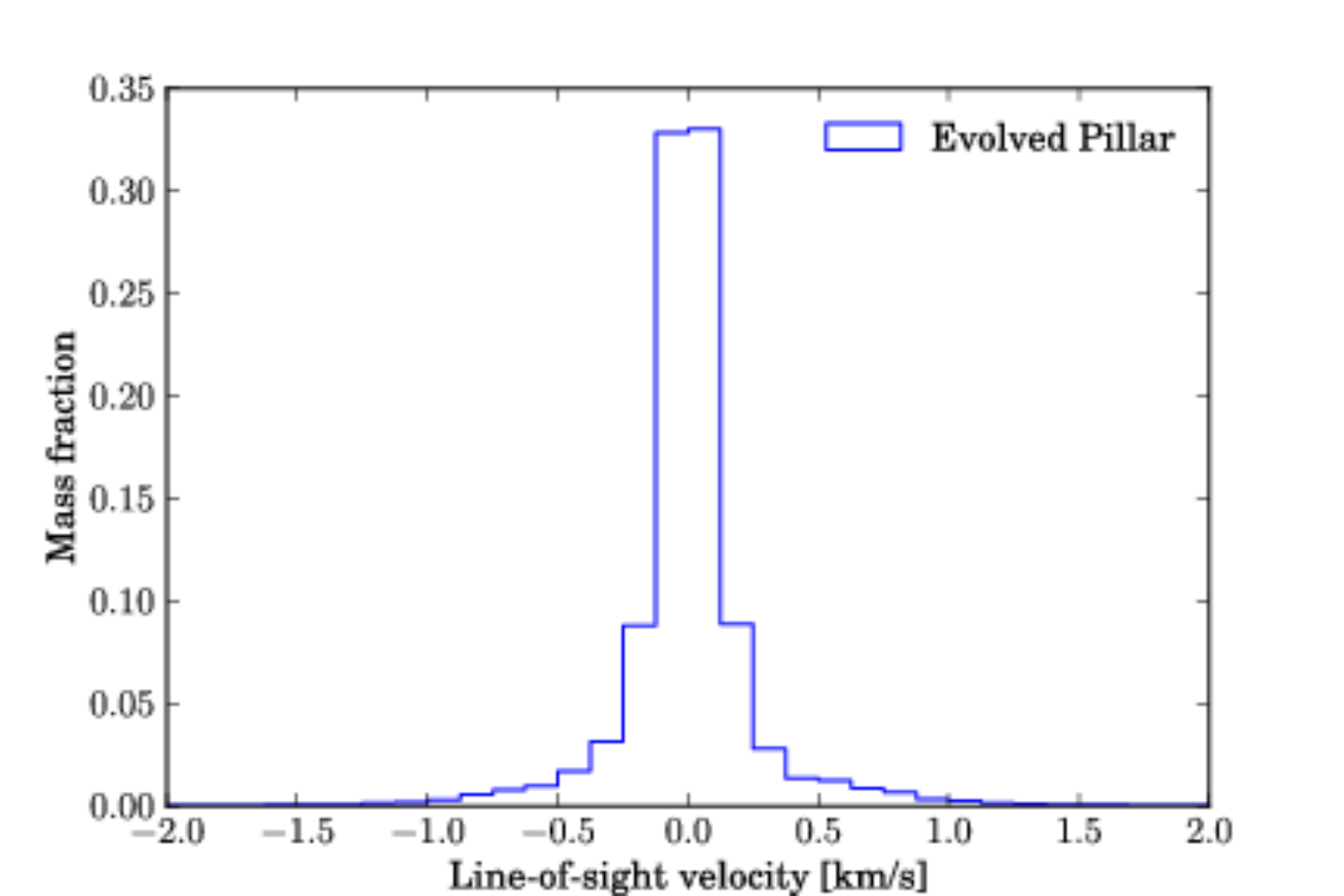}
\includegraphics[trim=0 0 0cm 0
  ,width=0.49\linewidth]{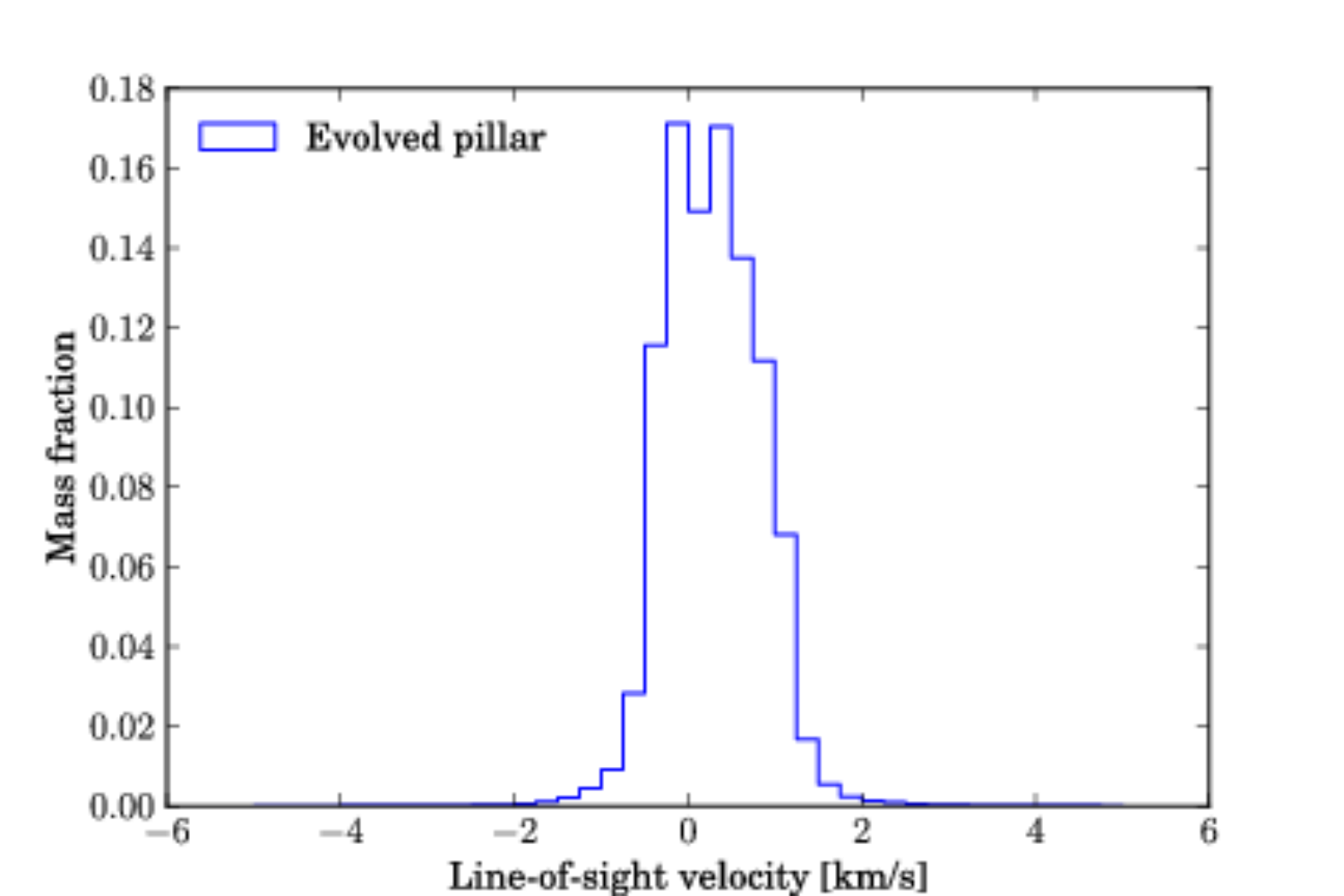} 

\caption{\label{simu_pillar} Distribution of the gas mass as a 
  function of the line-of-sight velocity for the snapshots in
  Fig.~\ref{simu_pillar_N}. Top: the spectra are 
  taken when the pillar is forming and the shell curved, just before the shell has
  collapsed in on itself. The blue- and red-shifted components can be
  clearly identified in both cases and are associated with the curved
  shell. Bottom: after the collapse, when the pillar 
  is in its growing state, the spectra display a single peak.} 
\end{figure}

With the DisPerSe algorithm, it is possible to detect pillars as
fronts with one side disconnected and pointing towards the ionizing
source and the other side connected at the base to the rest of the cavity. 
Long pillars (pillars of creation, the Spire, etc.) are detected well by
this process. This method is also useful for detecting small pillars at
the edge of the cavity, whic are possibly at the earliest phase of
their formation process. Such nascent pillars can be detected in the
Rosette and the Eagle nebula. In \citet{Tremblin:2012ej}, the key
ingredient for the formation of pillars is the high curvature of the
shell that collapses in on itself. A clear observational
diagnostic that has been derived is a red-shifted and a blue-shifted
  component that should appear in the velocity spectrum of the
nascent pillar as a signature of the collapse of the shell on itself
(see Figs.~\ref{simu_pillar} and~\ref{simu_pillar_N}, before the collapse). The same process
was also identified in turbulence simulations
\citep[see][]{Tremblin:2012he}.  This signature has
recently been confirmed by simulations with synthetic observations in $^{12}$CO and
$^{13}$CO performed by \citet{Haworth:2013gh}. All of these simulations
converge to highlight the collapse of the
shell on itself as the key process in forming a
pillar. Remarkably, the nascent pillar in the 
Rosette nebula detected by the DisPerSe algorithm presents such a
spectrum in $^{13}$CO (1-0). The same structure can also be seen in
the $^{12}$CO (1-0) data, as expected from the synthetic observations in
\citet{Haworth:2013gh}. Two central components at +13 and +14 km 
s$^{-1}$ are associated with the velocity components of the
front  east and west of the nascent pillar. In addition to
these central components, a red-shifted component is present at +19 km
s$^{-1}$ and a blue-shifted one at +11 km s$^{-1}$. The model in
\citet{Tremblin:2012ej} predicts that the velocity shift between the
two peaks is around twice the shell velocity if the line of
sight is aligned with the velocity of the collapsing fragments. If
not, the shell velocity is greater than half the velocity shift because
of the line-of-sight projection. The channel maps in
Fig~\ref{channel_maps} show that the 9-12 km/s and 17-21 km/s
components are not spatially aligned, so we expect a
projection effect on the velocity. 
Accordingly, we expect a
shell velocity higher than 4 km s$^{-1}$. If we assume it to be correct that the
shell velocity is measured by \citet{Kuchar:1993bn} at 4.5 km/s, we can estimate that
the velocities of the collapsing fragments have an angle of $\approx$ 25$^{\circ}$
with the line of sight.

In \citet{Tremblin:2012ej} and \citet{Tremblin:2012he}, evolved
pillars have a simpler velocity distribution, i.e., a single peak that
is centred on the velocity of the shell (see
Figs.~\ref{simu_pillar_N} and \ref{simu_pillar} ).
From the base, the velocity decreases along the pillar down to the
head. The gradient leads to a growth of the pillar as a function of
time over more than 1 My. More recent simulations (Tremblin et al. in
prep.) have shown that, at later times, the head is accelerated thanks to 
the rocket effect. Since the velocity at the base of the pillar
decreases as a function of time, the head can be accelerated to
velocities greater than the velocity of the base. At this point, the
pillar shrinks and can even be destroyed when the head reaches
the rest of the dense front. The spectra of the pillars in
the Rosette and Eagle nebula present a single-peak spectrum and a
gradient between the head and the base. Using the H$\alpha$
emission to examine the orientation
of the pillar, we are able to
conclude that the pillars are growing. Therefore they are evolved
pillars still in a phase of growth, and their heads are not accelerated
at velocities higher than the velocity of the shell.

\subsection{Turbulent globules}

\begin{figure}[t]
\centering
\includegraphics[trim=0 0 0cm 0
  ,width=0.8\linewidth]{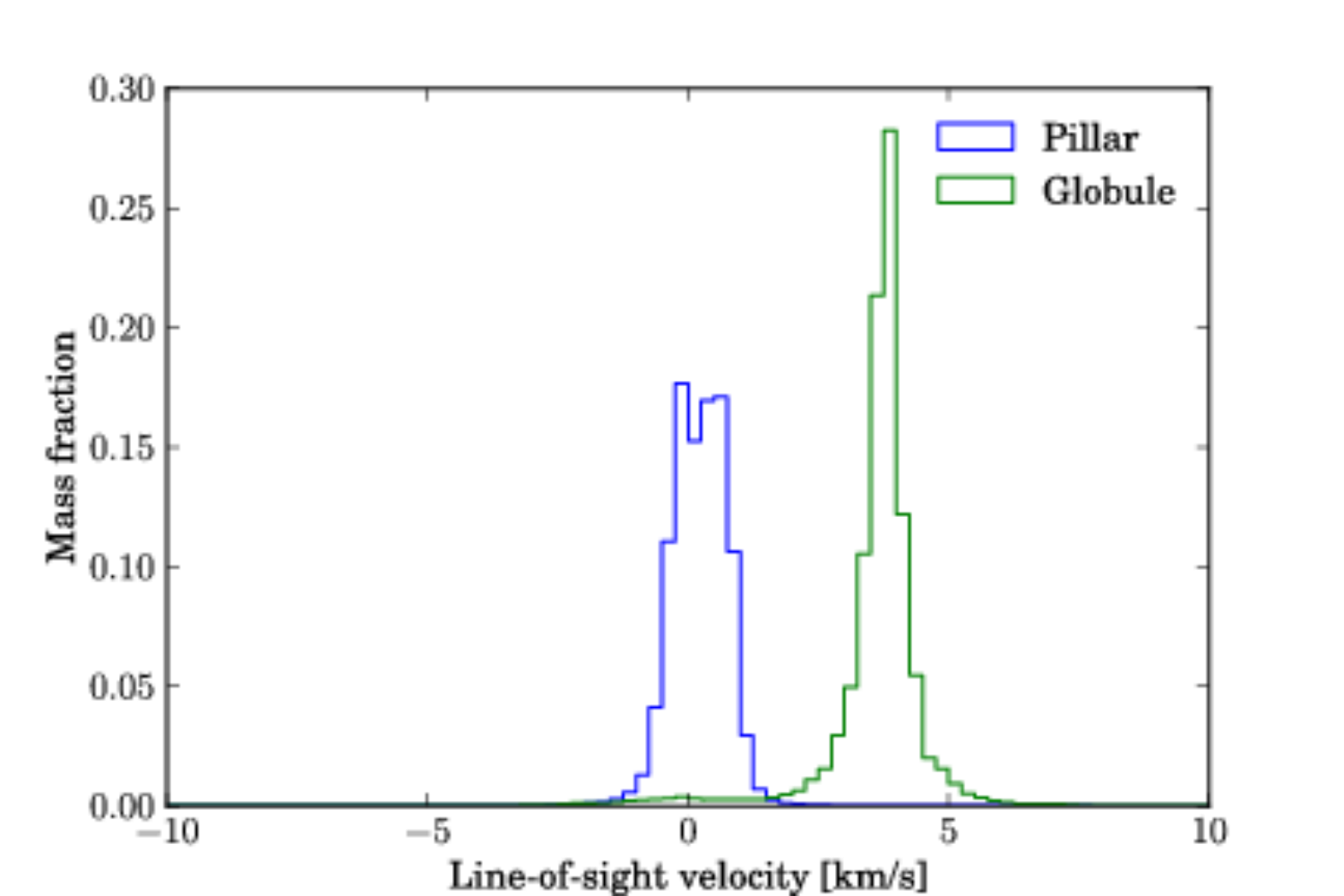} 

\caption{\label{simu_globule} Distribution of the gas mass as a
  function of the line-of-sight velocity for a pillar in
  the Mach-1 simulation and for a globule in the Mach-4 simulation
  \citep[see][for the details of the numerical
    setup.]{Tremblin:2012he}. Snapshots of the Mach 1 and Mach 4
  simulations are in  Fig.\ref{simu_turb_N}).} 
\end{figure}

\citet{Tremblin:2012he}
determined a link between the formation of globules in \ion{H}{ii}
regions and the initial turbulence of the molecular cloud. The process
is relatively simple: When the ram pressure of the turbulence is
greater than the pressure of the ionized gas, some parts of
cold gas have enough kinetic energy to penetrate the
\ion{H}{ii} region. Since their velocity is dominated by the
turbulence, they have a random component that is perpendicular to the
direction of expansion of the bubble, causing them to detach 
from the molecular cloud and form a cometary globule. Their transverse
random velocity can be used as observational diagnostics: pillars,
dense fronts, and condensations have a motion in the direction of
pressure-driven expansion of the bubble. In contrast, globules
will have a random motion that
will lead to large shifts in their bulk velocity compared to the
velocity of nearby dense fronts (see Fig.~\ref{simu_globule}). Such a
shift has already been observed in \citet{Schneider:2012hz} between a
globule and the nearby pillars at the edge of the ionized gas around
Cygnus OB2.

In the Rosette nebula, we observed two globules near
the dense front 2. The H$\alpha$ emission shows that this proximity is
not a projection effect.
The velocity spectra of these globules are shifted at + 3 km~s$^{-1}$
and + 6 km~s$^{-1}$ relative to the dense front, and since the structures are
close, these shifts can be interpreted as the relic of the initial turbulence.

\section{Conclusions}

Combining continuum and spectral line data, we are able to constrain the dynamics of the gas at the
edge of \ion{H}{ii} regions in the Rosette and Eagle nebulae. 

\begin{itemize}
\item The column-density dense fronts at the interface with the
  ionized gas present a compressed asymmetric profile with a width that is
consistent with numerical simulations whose key ingredients
  are turbulence and
ionization. 
\item A crest-detection algorithm is an efficient tool for
outlining pillars in the column density maps. The velocity structures of
these pillars are consistent with numerical simulations. Especially the detection
of a nascent pillar in the Rosette nebula was possible, and its
velocity profile presents blue-shifted and red-shifted components
in $^{12}$CO and $^{13}$CO. This type of profile was predicted in numerical
simulations as the signature of the collapse of the shell in on
itself, the first step in forming a pillar. The comparison of
the simulations and the observations in this paper confirms that this as a new scenario for the
formation of pillars. 
\item Existing numerical simulations show that globules tend to be
formed when the turbulent ram pressure is greater than the ionized-gas
pressure. When the turbulence dominates, some parts of the cold gas
have enough kinetic energy to penetrate the ionized region
and form cometary globules. Because of the link with turbulence, these
globules have a random bulk velocity set by the initial
turbulence, which can be perpendicular to the direction of the shell
expansion. Such dispersed bulk velocities are observed for the globules
in the Rosette nebula, thus confirming their turbulent origin. 
\end{itemize}
The observations presented in this paper converged with
previous numerical works to give a clear picture of the formation of
pillars and globules around \ion{H}{ii} regions, which are curved and
collapsing parts of the shells induced by pre-existing structures for
pillars and turbulent gas motion for globules. 
Gravity-based
scenarios, such as the Rayleigh Taylor instability first proposed by
\citet{Spitzer:1954jz}, or more recently, the remains of an accretion
flow \citep{Dale:2012gz}, are ruled out by the velocity
field structure in the case of the pillars of creation. Indeed,
\citet{Frieman:1954kj} showed that the effect of gravity would lead to
a free-fall profile for the velocity field; i.e., the velocity is
proportional to the square root of the distance along the pillar, and
\citet{Pound:1998hj} invalidated this possibility for the pillars of
creation. All
previous models do not predict a 
collapsing shell that is observed in the case of the nascent pillar in
the Rosette nebula. The structures both in M16 and Rosette are
consistent with the interaction between ionization and
turbulence, and other scenarios might still
happen in other regions, therefore more observations
are needed in order to confirm whether these two new models
are ubiquitous in galactic molecular clouds. 

A direct consequence of this pillar formation scenario is that a
pre-existing dense clump will curve the shell around it, triggering the
collapse and the formation of a pillar, while a low-density
perturbation will not curve the shell enough to trigger the
collapse. Instead, a dense clump is subsequently formed in the shell
by the accumulation of matter triggered by the perturbation. This
effect has been simulated in \citet{Tremblin:2012ej} and also in
\citet{Minier:2013ih} in the special case of RCW36. Therefore, star
formation happening at the tip of pillars is likely to have happened
even without ionization, while star formation in dense clumps in the
shell is likely to be triggered by the feedback \citep[see
  also][]{Dale:2013da}. With the recent 
statistics of YSOs located at the edge of ionized bubbles
\citep[e.g.][]{Thompson:2012gn}, it becomes clear that star formation
triggered by the feedback of massive stars is indeed frequent and an
important ingredient for understanding galactic star formation.

%
%

\begin{acknowledgements}
SPIRE has been developed by a consortium of institutes led by Cardiff
Univ. (UK) and including: Univ. Lethbridge (Canada); NAOC (China);
CEA, LAM (France); IFSI, Univ. Padua (Italy); IAC (Spain); Stockholm
Observatory (Sweden); Imperial College London, RAL, UCL-MSSL, UKATC,
Univ. Sussex (UK); and Caltech, JPL, NHSC, Univ. Colorado (USA). This
development has been supported by national funding agencies: CSA
(Canada); NAOC (China); CEA, CNES, CNRS (France); ASI (Italy); MCINN
(Spain); SNSB (Sweden); STFC, UKSA (UK); and NASA (USA). 
PACS has been developed by a consortium of institutes led by MPE
(Germany) and including UVIE (Austria); KU Leuven, CSL, IMEC
(Belgium); CEA, LAM (France); MPIA (Germany); INAF-IFSI/OAA/OAP/OAT,
LENS, SISSA (Italy); IAC (Spain). This development has been supported
by the funding agencies BMVIT (Austria), ESA-PRODEX (Belgium),
CEA/CNES (France), DLR (Germany), ASI/INAF (Italy), and CICYT/MCYT
(Spain). Part of this work was supported by the  ANR-11-BS56-010  
project ``STARFICH''.  
\end{acknowledgements}

%
%

\bibliographystyle{aa}
\bibliography{main.bib}

\end {document}